\documentclass[lettersize,journal]{IEEEtran}
\usepackage{graphicx}
\usepackage{setspace}
\usepackage{adjustbox} 
\DeclareGraphicsExtensions{.pdf,.jpeg,.png}
\usepackage{enumitem}
\usepackage[linesnumbered,ruled]{algorithm2e}
\usepackage{amsmath}
\usepackage{lipsum}
\usepackage{mathtools}
\usepackage{cuted}
\usepackage{array}
\usepackage{subfigure}
\usepackage{amssymb}
\usepackage{caption}
\usepackage{multirow}
\usepackage{makecell}
\usepackage{algorithmic}
\usepackage{xcolor}

\SetCommentSty{mycommfont}
\usepackage{array,multirow,makecell}
\SetKwInput{KwInput}{Input}    
\SetKwInput{KwRepeat}{Repeat} 
\SetKwInput{KwOutput}{Output}              
\begin{document}
\bstctlcite{IEEEexample:BSTcontrol}
    \title{\huge Optimizing Multi-User Uplink Cooperative Rate-Splitting Multiple Access: Efficient User Pairing and Resource Allocation }
\author{Shreya Khisa, Mohamad Elhattab, Chadi Assi and Sanaa Sharafeddine \vspace{-0.4cm}}
\maketitle

\begin{abstract}
This paper investigates joint user pairing, power and time slot duration allocation in the uplink multiple-input single-output (MISO) multi-user cooperative rate-splitting multiple access (C-RSMA) networks in half-duplex (HD) mode. We assume two types of users: cell-center users (CCU) and cell-edge users (CEU); first, we propose a user pairing scheme utilizing a semi-orthogonal user selection (SUS) and a matching-game (MG)-based approach where the SUS algorithm is used to select CCU in each pair which assists in reducing inter-pair interference (IPI). Afterward, the CEU in each pair is selected by considering the highest channel gain between CCU and CEU. After pairing is performed, the communication takes place in two phases: in the first phase, in a given pair, CEUs broadcast their signal, which is received by the base station (BS) and CCUs. In the second phase, in a given pair, the CCU decodes the signal from its paired CEU, superimposes its own signal, and transmits it to the BS. We formulate a joint optimization problem in order to maximize the sum rate subject to the constraints of the power budget of the user equipment (UE) and Quality of Service (QoS) requirements at each UE. Since the formulated optimization problem is non-convex, we adopt a bi-level optimization to make the problem tractable. We decompose the original problem into two sub-problems: the user pairing sub-problem and the resource allocation sub-problem where user pairing sub-problem is independent of resource allocation sub-problem and once pairs are identified, resource allocation sub-problem is solved for a given pair. Resource allocation sub-problem is solved by invoking a successive convex approximation (SCA)-based approach. Simulation results demonstrate that the proposed SUS-MG-based algorithm with SCA outperforms other conventional schemes. 
\end{abstract}

\IEEEpeerreviewmaketitle
\section{Introduction}
The significant increase in wireless traffic and the growing demand for high-speed data transmission have sparked considerable interest in innovative solutions aimed at advancing the upcoming phase of wireless communication, often referred to as the sixth generation, (6G) \cite{9349624}. In the evolution of 6G, it is imperative to address the escalating need for ultra-high reliability, high throughput, diverse Quality of Service (QoS) requirements, ultra-low latency, and massive connectivity. These factors are pivotal in fulfilling the requirements of services such as extremely reliable and low-latency communication (eURLLC), enhanced mobile broadband (eMBB), and ultra-massive machine type communication (umMTC) \cite{8869705}. These challenges are compounded by the rapid proliferation and widespread utilization of smartphones and tablets. The increasing number of these devices will inevitably congest the wireless spectrum further, exacerbating the scarcity of available spectrum resources. To combat the spectrum crunch and satisfy the rigorous demands of surging broadband usage, one effective approach is to explore innovative and efficient multiple access (MA) technologies, which have the potential to enhance system capacity in a cost-effective manner.
\par Recently, rate-splitting multiple access (RSMA) has emerged as a promising contender for a non-orthogonal MA mechanism, offering flexible interference management for next-generation wireless communication \cite{9831440}. The main principle behind RSMA is to partially treat the multi-user interference as noise and partially decode it \cite{9831440}. Utilizing the RSMA principle, the base station (BS) splits the signals of user equipments (UEs) into common and private parts and transmits the total signal using superposition coding (SC). At the receiver side, the common stream is decoded by treating all private streams as interference, and then, the decoded common stream is removed from the total received signal utilizing successive interference cancellation (SIC) process. Meanwhile, the private streams are decoded at a particular UE by treating other private streams coming from other UEs as interference. It should be noted that the common streams are needed to be decoded by all users. On the other hand, private streams are decoded by their intended users only. This flexible nature of the interference management scheme assists RSMA to bridge the gap between space-division multiple access (SDMA), which fully treats multi-user interference as noise, and non-orthogonal multiple access (NOMA), which fully decodes the interference \cite{9831440}.
\par Even though RSMA has shown significant improvement in performance gain over NOMA and SDMA in terms of throughput, sum-rate, and energy efficiency \cite{9831440}, it may suffer from performance loss, which may limit its potential gain. This is because the common stream is required to be decoded by all users, and hence, the achievable common rate is constrained by the worst-case user who possesses a poor channel gain with the BS. In order to tackle this challenge and unleash the full potential gains of RSMA, the amalgamation between cooperative communication and RSMA has been investigated, which is known as cooperative RSMA (C-RSMA) \cite{9771468}, \cite{9852986}, \cite{9627180}. Specifically, in C-RSMA, the cell-center users (CCUs), which maintain a good channel gain with the BS, can assist the cell-edge users (CEU)s by relaying the decoded common stream to the CEUs to improve their signal quality. Consequently, C-RSMA has shown promising results in terms of rate region \cite{8846761}, user fairness \cite{9123680}, \cite{9771468}, power consumption minimization \cite{9627180}, network coverage extension \cite{9831440}, and secrecy rate enhancement \cite{9831440} in comparison to the traditional RSMA.
\par It should be noted that all the above-mentioned works mainly focused on the C-RSMA framework in a downlink scenario, meanwhile C-RSMA framework in the uplink setup is still in its infancy stage, which motivates this study. The primary difference between uplink RSMA and downlink RSMA is in splitting the transmitted signal for each user \cite{9257190}. Specifically, in uplink, UEs can split their signals into multiple parts without considering any common part or private part. It implies that there is no common message transmission in the uplink RSMA scenario. {Particularly, according to the principle of uplink RSMA, at user-$k$, $k \in \{1,..., K\}$, the message $W_k$
to be transmitted is split into two sub-messages $W_{k,1}$ and
$W_{k,2}$. This can be interpreted as creating two virtual users \cite{9831440}. The messages $W_{k,1}$ and $W_{k,2}$ of the two virtual users are
independently encoded into streams $s_{k,1}$ and $s_{k,2}$ with unit
variance, i.e., $E[|s_{k,i}|^2]=1, i = 1, 2$. These two streams are then respectively allocated with certain powers, $P_{k,1}$ and $P_{k,2}$, and superposed at user-$k$.  The main advantage of uplink RSMA over uplink NOMA lies in the flexible decoding process at the BS. Particularly, sub-messages of RSMA belonging to a particular user do not need to be decoded sequentially, and the decoding of sub-messages totally depends on the adopted decoding order. For example, sub-message 1 of user 2 can be decoded before the sub-message 2 of user 1. This flexible decoding nature of RSMA helps to decode one sub-message of a particular user with more interference and another sub-message of the same user can be decoded with less interference. On the other hand, in NOMA, as no message is split, the whole message of a particular user is decoded at the BS while considering the other user messages as interference, resulting in non-flexible interference management.} Even though uplink RSMA may suffer from a higher number of  SIC processes, however, SIC occurs at BS, which has high processing capabilities. In addition, in uplink RSMA, the rate of each user is the summation of split messages whereas the rate of each
user in uplink NOMA comes from decoding a single message. Hence, uplink RSMA is able to achieve a higher rate than uplink NOMA. Motivated by the above-mentioned benefits, this paper investigates the integration of uplink RSMA with cooperation in a multi-user scenario where two users are paired in such a way that it maximizes the overall sum-rate of the system.
\begin{table*} \label{table1}
\captionof {table}{Summary of key symbols}
\begin{adjustbox}{max width=\textwidth}
\centering
\begin{tabular}{ |c|l| }
 \hline
  \textbf{Symbol} & \textbf{Description}  \\
 \hline
 $N$ & Number of antennas  at the BS.  \\
 \hline
 $\mathcal{K}, \mathcal{U}, \mathcal{V}$ & \makecell[l]{Set of all UEs, set of all CCUs, and set of all CEUs, respectively. }\\
 \hline
\makecell{$\textbf{h}_{u},\textbf{h}_{v},{h}_{v,u}$} & {\makecell[l]{Channel coefficient of BS $\rightarrow$ CCU-$u$, BS $\rightarrow$ CEU-$v$, and CCU-$u$ $\rightarrow$ CEU-$v$, respectively.}}\\
 \hline
$\mathcal{B}$, $b$& \makecell[l]{Set for the number of sub-messages, and index for sub-message, respectively.} \\
\hline
$s_{u,b}$ & Sub-message with index $b$ transmitted by CCU-$u$.\\
\hline $P_{u,b}$ & \makecell[l]{The power for CCU-$u$ to transmit sub-message $s_{u,b}$}.\\
\hline
$\boldsymbol{y}_{BS}^{[1]}$, $\boldsymbol{y}_{BS}^{[2]}, y_{v\rightarrow u}^{[1]}$ & \makecell[l]{The signal received at the BS in the DT phase, the signal received at the BS in the \\ CT phase, and signal received at CCU-$u$ due to the transmission of CEU-$v$ in the DT phase, respectively.}\\
 \hline
 $\delta$, $\theta$ & Time slot duration, SUS factor.  \\
\hline
$I_{v' \rightarrow u}$, $I_{u',BS}$, $I_{v',BS}$ &  \makecell[l]{The interference received at CCU-$u$ due to the transmissions of all CEUs except its paired CEU-$v$, at the DT phase, to \\ recover
a signal of a  particular CCU-$u$ at the BS, the interference that is calculated at BS, originating from all CCUs\\ except the corresponding CCU-$u$ at CT phase, to recover a signal of a particular CEU-$v$ at BS, interference that is\\ calculated at the BS, originating from all CEUs except the corresponding CEU-$v$ at the DT phase, respectively. }\\
 \hline
 $P_u^{max}$, $P_v^{max}$ & The power budget of each CCU-$u$ and the power budget of each CCU-$v$, respectively.\\
 \hline
 $R_{th,k}$ & The QoS constraints in terms of the minimum required data rate for user-$k$.\\
 \hline
 $\boldsymbol{n}_v$, $\boldsymbol{n}_u$, ${n}_{v,u}$ & The Additive white Gaussian noise (AWGN) with $\mathcal{CN}\left(0, 1\right)$. \\
 \hline
\end{tabular}
\end{adjustbox}
\label{table1}
\end{table*}
\subsection{State of the Art}
The investigation regarding the performance of rate splitting (RS) has started from the perspective of information theory which has shown to achieve the
optimal sum Degree of Freedom (DoF) \cite{7555358}. Afterward, this investigation is followed by the performance evaluation of RS in multiple-input and single-output (MISO) broadcast (BC) scenarios with imperfect Channel State Information at the Transmitter (CSIT) \cite{7972900}. Following this, several attempts have been made to evaluate the performance of RSMA in different networks and integrate RSMA with different advanced technologies. For example, the authors in \cite{mao2018rate} studied the performance of RSMA in different scenarios of overloaded and underloaded networks. It also showed the performance gain that RSMA can achieve over conventional NOMA, SDMA, and OMA. It is one of the early investigations on RSMA in the downlink network. The authors in \cite{9461768} studied the sum-rate maximization problem for wireless networks in downlink RSMA. The authors in \cite{9491092} investigated the performance of RSMA under the imperfect CSIT due to user mobility and latency/delay in the network. Besides the conventional MISO BC framework, the advantages of RS have been further explored in satellite communications \cite{9324793}, Cloud Radio Access Network (C-RAN) \cite{9445019},  massive multiple-input and multiple-output MIMO \cite{9491092}, reconfigurable intelligent surface (RIS) \cite{9847599}, radar communications \cite{9531484}, multi-cell coordinated multipoint joint transmission (CoMP) \cite{8756668}, and simultaneous wireless information and power transfer (SWIPT) \cite{9849099}.
\par Recently, several studies have been carried out to show the performance gain of C-RSMA in both full-duplex (FD) and half-duplex (HD) modes. The authors in \cite{9771468} studied the performance of C-RSMA in FD mode for two users' cases. The authors in \cite{9123680} investigated the C-RSMA framework for $K$-users, where each CCU relays the common stream to all CEUs in HD mode. In \cite{9627180}, an FD C-RSMA scheme in a downlink two-group multicast system was studied where CCUs can harvest energy from BS using SWIPT, and then utilize this harvested energy to relay the common stream to CEUs. It should be noted that all of the above-mentioned works studied the RSMA and C-RSMA schemes in the downlink framework. 
\par On the other hand, several works have studied the performance of RSMA in uplink. The authors in \cite{9257190} investigated the sum rate maximization problem of RSMA in an uplink network. The authors in \cite{10190330} analyzed the performances of different network slicing schemes in uplink based on RSMA.
The authors in \cite{9970313} investigated the sum throughput and error
probability of uplink RSMA with fixed block length (FBL) coding
in a two-user system. Meanwhile, authors in \cite{9676684} studied the
performance of an uplink RSMA
network with two sources, in terms of outage probability and
throughput. The authors in \cite{10167483} investigated
the outage performance of uplink RSMA transmission with
randomly deployed users, taking both users
scheduling schemes and power allocation strategies into consideration. An RIS assisted uplink RSMA system \cite{9912342} is investigated for dead-zone users where the direct link
between the users and the BS is unavailable. The authors in \cite{9991090} investigated the user fairness of downlink multi-antenna RSMA in short-packet
communications with/without cooperative (user-relaying) transmission.
In addition, RSMA in uplink has been investigated in satellite communications \cite{9790069}, RIS-assisted wireless networks \cite{9912342}, \cite{10102273}, \cite{10014691}, massive MIMO \cite{10032157}, integrated sensing and communications \cite{10119024}, unmanned aerial vehicle (UAV)-assisted networks \cite{10059199} and so on. However, the uplink RSMA in cooperative communications, i.e., C-RSMA, is still in its development stage.
\par Recently, the authors in \cite{9852986} demonstrated the effectiveness of C-RSMA and cooperative NOMA in the uplink framework where both users cooperate with each other to relay each other signals to the BS. However, this work studied only a simple two-user case scenario, and a single antenna BS was considered. Hence, it lacks considerable challenges resulting from the multi-user and multi-antenna BS settings, such as multi-user interference and designing beamforming vectors at BS. Furthermore, it is worth mentioning that this study did not take into account any pairing methods, which is vital for maximizing the advantages of cooperative communication. Motivated by this fact, to the best of our knowledge, this is the first paper that considers the C-RSMA framework in a multi-user scenario in uplink communication. In this work, we investigate the C-RSMA framework in a multi-user scenario by proposing a novel pairing policy, optimizing the power allocation of the UEs, and time slot duration allocation for communication and cooperation while minimizing inter-pair interference (IPI).
\subsection{Contributions}
To the best of our knowledge, the study of C-RSMA in a multi-user uplink network scenario has not been explored to date. To fill this research gap, we propose a semi-orthogonal user selection and a matching game (SUS-MG)-based sum rate maximization problem for uplink MISO C-RSMA framework. The main contributions of this paper are outlined as follows.
\begin{itemize}
    \item We formulate an optimization problem to maximize the sum rate of the uplink MISO C-RSMA framework by jointly optimizing user pairing and resource allocation. The formulated optimization problem results in a mixed-integer non-linear problem (MINLP). Hence, we solve the problem by invoking bi-level optimization. Specifically, we decompose the original problem into two sub-problems: the user pairing sub-problem and the resource allocation sub-problem. 
    \item In the user pairing sub-problem, we propose a semi-orthogonal user selection (SUS) \cite{9110852} and matching game (MG)-based user pairing scheme \cite{9737471} that can suppress IPI and create efficient pairs to maximize the system's performance. Specifically, utilizing the SUS algorithm, we determine the CCU in each pair by reducing IPI. Afterward, an MG-based strategy is used to determine the CEU in each pair considering the highest channel gains between all CCUs and corresponding CEU as a utility function. 
    \item In the resource allocation sub-problem, we employ a low-complexity successive convex approximation (SCA)-based algorithm to optimize the transmit power of each relaying UE and the allocation of time slots for communication with the BS and UE cooperation within a given pair. In this framework, the communication takes place in two transmission phases: the direct transmission (DT) phase and the cooperative transmission (CT) phase. During the DT phase, all CEUs broadcast their signals and they are received at the BS and CCUs. Meanwhile, during the CT phase, each CCU transmits the decoded signal of its paired CEU to the BS and also superimposes its own signal. In addition, we utilize maximum ratio combination (MRC) equalization to recover the signal at the BS by considering the IPI. 
    \item {Through an extensive experiment, we evaluate the impact of splitting messages in CCUs and CEUs where we have demonstrated that splitting one user message in each pair is enough to achieve better performance than splitting both user messages in each pair. }
\end{itemize}
Through experiments, we have demonstrated the effects of the different decoding orders on the average sum rate in the uplink C-RSMA framework. From extensive simulations, we have found a best-performing decoding order for uplink MISO C-RSMA. Finally, our simulation results demonstrated that the proposed SUS-MG-SCA algorithm achieves higher performance over random pairing and other conventional MA schemes for different values of power budget constraints at both CCUs and CEUs, and QoS constraints at each UE.
\subsection{Paper Organization and Notations}
The rest of the paper is organized as follows. Section \ref{System Model} presents the system model. Section \ref{SINRss} presents the decoding order and the achievable data rate analysis. Section \ref{Problem Formulation} discusses the formulated optimization problem and the solution roadmap. Meanwhile, Section \ref{pairing} and \ref{SCA} provide the details of the proposed solution approach.  Finally, the simulation results and the conclusion are discussed in Sections \ref{Simulation} and \ref{Conclusion}, respectively. A summary of key symbols is provided in Table \ref{table1}. Matrices and vectors are denoted by bold-face lower-case and upper-case letters, respectively. For any complex-valued vector $x$, $||x||$ refers to the norm of vector $x$, $(.)^H$ represents Hermitian transpose, $E\{\}$ is the expectation operator of a random variable, and $\mathcal{R}\{x\}$ is the real part of the complex term $x$.
\section{System Model}
\label{System Model}
\subsection{Network Model}
\begin{figure}[!t]
    \centering
    \includegraphics[width=0.7\columnwidth]{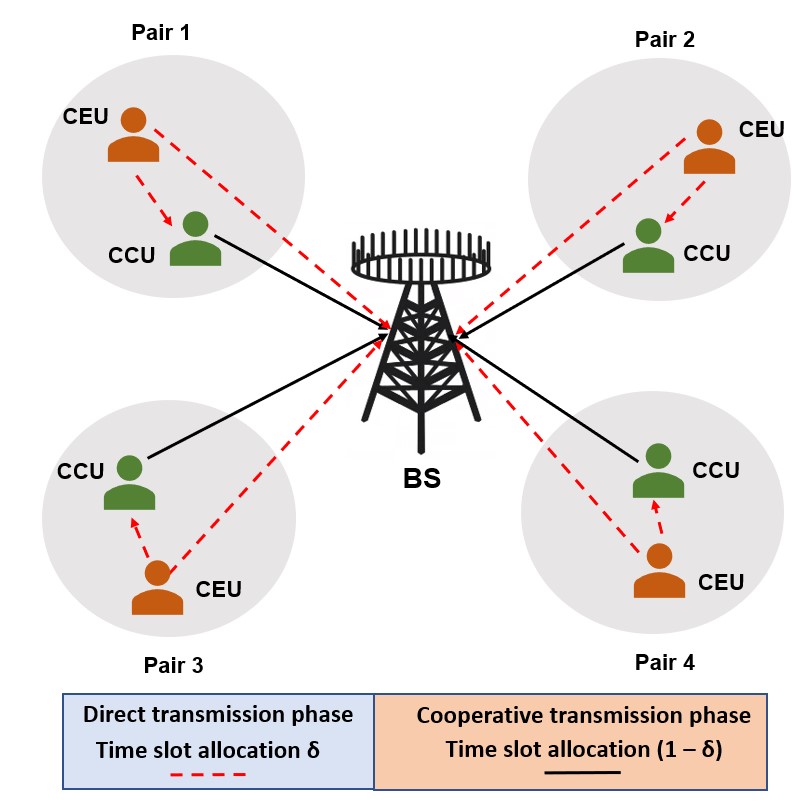}
    \caption{Uplink C-RSMA: $K=8$ users, 4 pairs}
\label{Fig 1: System Model}
\end{figure}
We consider an uplink transmission of a single-cell C-RSMA system consisting of one BS with $N$ antennas, $K$ single-antenna UEs where $\mathcal{K}=[1,2,3,\dots,K]$ as shown in Fig. 1. The BS serves $K$ users in the same frequency-time resource block. Additionally, the RSMA technique is invoked as the MA scheme, which enables multiple users to share the same channel resources in order to enhance the system's spectral efficiency. Two types of UEs are considered for this model: CCUs and CEUs. Specifically, CCUs sustain a good channel condition with the BS, and the set of all CCUs are denoted by $\mathcal{U}={[1,2,\dots,U]}$, where $U=\frac{K}{2}$. On the other hand, CEUs experience poor channel conditions with the BS, and the set of all CEUs can be denoted as  $\mathcal{V}={[1,2,\dots, V]}$, where $V=\frac{K}{2}$. Note that the CCUs are capable of acting as HD relays and can forward the messages of the CEUs to the BS. Therefore, one CCU and one CEU should be paired in an efficient way so that the CCU is able to assist in improving the signal quality of its paired CEU. A pair of CCU and CEU can be denoted as $(u,v)$, $\forall u \in \mathcal{U}$, and $\forall v \in \mathcal{V}$.
%
%
\par We denote the wireless channel link between BS and each CCU-$u$ as $\textbf{h}_{u} = [h_{u,1}, h_{u,2},\dots,h_{u,N}]^T \in \mathbb{C}^{N \times 1}$, where $h_{u,n}$ represents the channel response of the wireless link from the CCU-$u$ to the $n$-th antenna element of the BS. Meanwhile, the wireless channel link between the BS and each CEU can be denoted as $\textbf{h}_{v}= [h_{v,1}, h_{v,2},\dots,h_{v,N}]  \in \mathbb{C}^{N \times 1}$, respectively. Note that each link $h_{k,n}$ is represented as a product of path loss factor, $\tau$, and small scale fading, $\xi$. 
Particularly, each channel response between $n$-th antenna element and UE-$k$ can be represented as $h_{k,n}= \sqrt{\tau_{k,n}}\xi_{k,n}$. Since the antenna elements are located in very close proximity to each other, without loss of generality, the path loss factor for the same UE is assumed to be equal. Specifically, $\tau_{k,1}=\tau_{k,2}=\tau_{k,3}=\dots \triangleq \tau_{k,N}$. We assume that the small-scale fading factor $\xi_{k,n}$ is independent and identically distributed such that $E\{|\xi_{k,n}|^2\}=1,\forall k \in \mathcal{K}, \forall n \in N$. In addition, in our model, we consider IPI in our signal transmission which results from the interference coming from members from other pairs except the corresponding pair. 
\subsection {Transmission model}
The whole communication takes place in two-tim slots. The first slot is referred to as the direct transmission (DT) phase, and the second slot is called the cooperative transmission (CT) phase. As illustrated in Fig \ref{Fig 1: System Model}, the time slot allocation for the two phases may not be equal. DT phase occurs in $\delta$T time duration. Meanwhile, CT phase occurs in $(1-\delta)$T duration. A detailed description of the two phases is provided below.\\
\textbf{1.\textit{ Direct transmission (DT) phase}}: Here, {CEUs broadcast their signals which are received by the BS and CCUs.} Moreover, the signal of CEUs suffers from IPI at the BS. Particularly, the signal of each CEU is interfered {by} the signals of all other CEUs in the system as we assume that all CEUs are transmitting at the same time. After receiving the signal from its paired CEU, each CCU decodes the signal of its paired CEU.\\
\textbf{2. \textit{Cooperative transmission (CT) phase}}: During this phase, we utilize non-regenerative decode-and-forward (NDF) protocol. {Specifically, the definition of NDF
protocol states that after the relay node receives the signal from the source node, it decodes the
received signal. However, it re-encodes the signal with a codebook generated independently
from that of the source node and transmits it in the second channel of the source node \cite{9123680,5502106,8846761}.
It is important to note that in the NDF protocol, a time frame of $T$ seconds is divided into two-time slots: $\delta T$, and $(1-\delta)T$.} Hence, utilizing NDF protocol, each CCU re-encodes the decoded message with a codebook that is different from its paired CEU. Afterward, each CCU superimposes its own signal and forwards the total signal to the BS. However, the signal of each CCU interferes with the signals of other CCUs located in different pairs resulting in IPI. 
\subsection{Signal Model}
{By utilizing the uplink RSMA principle, the messages of all CCUs are split into two sub-messages, and the set of sub-messages is denoted as $\mathcal{B}=[1,2]$. However, the messages of CEUs are kept without splitting.  Afterward, all the sub-messages of CCUs and messages of CEUs are encoded independently, and hence, generate streams $s_{u,b}$, $\forall u \in \mathcal{U}, \forall b \in \mathcal{B}$ and $s_{v}, \forall v \in \mathcal{V}$. For simplicity, we analyze the signal model of a single pair $(u,v)$, which is provided below:}\\
\textbf{1. \textit{Signal transmission during DT phase}}: During the DT phase, the CEU-$v$ broadcasts its signal which is received by the both BS and CCU-$u$. The signal received by the BS due to the transmissions of CEU-$v$, $\forall v \in \mathcal{V}$ at time slot $\delta T$ can be expressed as,
\begin{equation}
\begin{split}
   {\boldsymbol{y}_{BS}^{[1]}=\sum_{v \in \mathcal{V}}\left(\textbf{h}_{v}\sqrt{P_{v}}s_{v}\right)+ 
    \boldsymbol{n}_{v}}.\label{Sig1}
    \end{split}
\end{equation}
Meanwhile, the signal that is received by CCU-$u$ due to the transmission of a CEU-$v$ at time slot $\delta T$ can be presented by,
\begin{equation}
\begin{split}
{
y_{v\rightarrow u}^{[1]}=h_{v,u}\sqrt{P_{v}}s_{v}+I_{v' \rightarrow u}+{n}_{v,u}, \label{Sig2}}
\end{split}
\end{equation}
where {$I_{v' \rightarrow u}=\sum_{v' \in \mathcal{V}, v' \neq v}h_{v',u}\sqrt{P_{v'}}s_{v'}$.} \\
\textbf{2. \textit{Signal transmission during CT phase}}: During the CT phase, which is denoted by time slot $(1-\delta)T$, utilizing NDF protocol, the CCU-$u$ re-encodes the decoded signal utilizing a different codebook than the original one. Afterward, CCU-$u$ superimposes its own signal and forwards the superimposed signal to the BS. Hence, after the CT phase, the signal received by the BS can be given by, 
\begin{equation}\footnotesize
{\boldsymbol{y}_{BS}^{[2]}=
\sum_{u \in \mathcal{U}}\textbf{h}_{u}\left(\sum_{b \in \mathcal{B}}\left(\sqrt{P_{u,b}}s_{{u,b}}\right)+\sqrt{P_{u,\hat{v}}}s_{u,\hat{v}}\right)
+\boldsymbol{{n}}_{u},} \label{Sig3}
\end{equation}
where $P_{u,\hat{v}}$ represents the transmission power of CCU-$u$ to transmit the message of CEU-$v$ that is received during the DT phase.
\subsection{Signal recovery at BS}
The BS receives signals from all $K$ users at the end of the two transmission phases. In order to recover each signal at the BS, at first, we invoke the MRC equalization to separate the signals of different UEs. Afterward, the SIC is utilized to remove the decoded signal from the total received signal. Details about the recovery process are provided below:\\
\textbf{1. \textit{MRC Equalization}}: In the first step of the recovery process, the MRC equalization is utilized in order to separate signals of different UEs. In this process, the received signal is multiplied by the conjugate transpose of the CSI. \footnote{We assume perfect channel state information (CSI) at the BS. Therefore, the BS perfectly knows every wireless link to any UE \cite{9123680, liu2022effective}.} For example, we multiply $\textbf{h}_{u}^H$ which is associated with the CCU-$u$ with the total received signal at the CT phase. After MRC operation, the restored message of CCU-$u$ can be denoted by, 
\begin{equation}
    \hat{s}_{u,b}=\frac{1}{N}\textbf{h}_{u}^H\left(\boldsymbol{y}_{BS}^{[2]}\right), \label{Sig5}
\end{equation}
\begin{equation}\footnotesize
\begin{split}
{
    \hat{s}_{u,b}= \frac{||\textbf{h}_{u}||^2 }{N}\left(\sum_{b \in \mathcal{B}}\left(\sqrt{P_{u,b}}s_{u,b}\right)+\sqrt{P_{u,\hat{v}}}s_{u,\hat{v}}\right)+I_{u',BS}+\frac{\textbf{h}_u^H}{N}\boldsymbol{{n}}_u, }\label{Sig6}
    \end{split}
\end{equation}
where \begin{equation}\footnotesize
\begin{split}
& {I_{u',BS} =}\\
& {\sum_{u' \in \mathcal{U}, u' \neq u}\left(\frac{\textbf{h}_u^H\textbf{h}_{u'}}{N}\sum_{b \in \mathcal{B}}\left(\sqrt{P_{u',b}}s_{u',b}\right)+\sqrt{P_{u',\hat{v}}}s_{u',\hat{v}}\right).} \notag
\end{split}
\end{equation}
\par Note that the MRC receiver maximizes the user’s signal power without excessively boosting the noise. The MRC can achieve near-optimal performance with a massive number of antennas \cite{5595728}. Moreover, the MRC attains much lower complexity in comparison to linear minimum mean square error (LMMSE) and zero-forcing (ZF) equalizers \cite{5595728}, because it does not involve the calculation of the matrix inverse \cite{5595728}. Hence, the MRC becomes the most efficient linear equalizer as long as sufficient degrees of freedom can be leveraged by a large number of antennas \cite{8472883}. It should be noted that when the BS contains a large number of antennas, $\frac{||\textbf{h}_k||^2}{N}$ is close to $\tau_{k,n}$, i.e. $\frac{1}{N}\textbf{h}_k^H\textbf{h}_k \xrightarrow{a.s} \tau_{k,n}$ where $\xrightarrow{a.s}$ denotes the almost sure convergence \cite{6457363}. This phenomenon is called channel hardening \cite{6457363}. The channel hardening refers to the situation when
the randomness of the channel fading coefficients decreases due to the existence of large antenna arrays \cite{9069181}. Hence, the received useful signal $s_{k,b}$ is scaled by a real-valued coefficient $P_{k,b}\tau_{k,b}$. Meanwhile, when the number of antennas is very large, $N\rightarrow \infty$,  ${\frac{1}{N}\textbf{h}_k^H\textbf{h}_{k'}} \xrightarrow{a.s} 0$. This phenomenon implies that the channel responses of different users tend to be quasi-orthogonal among each other when the number of BS antennas is large enough. Hence, the interference coming from other users is reduced \cite{7294693}. However, in a real-life practical scenario, it is not always possible to have a very high number of (infinite) antennas and it is not possible to completely diminish the interference coming from different users. Considering this, in our work, we have taken the effect of the IPI into consideration. 
\par\textbf{2. \textit{SIC Process and Decoding Order}}: After we separate the signal of a particular UE using MRC equalization, the sub-messages/ messages of that UE are decoded based on their decoding order such that the users with low decoding order are decoded first and the users with high decoding order are decoded later. This process is provided in detail in the following section.
\vspace{-0.3cm}
\section{Decoding Order and Achievable Rate Analysis}
\label{SINRss}
\subsection{Decoding Order}
{The decoding order is an important factor in decoding the received signals during the CT phase. Assuming that the decoding orders of sub-messages/messages at the BS are denoted as the set $\pi_{BS} = [\pi_{u,b}, \pi_{u,\hat{v}}, \pi_{v}| \forall u \in \mathcal{U}, \forall b \in \mathcal{B}, \forall v \in \mathcal{V}]$ respectively, and the messages/sub-messages are decoded in the ascending order. $\pi_{u,b}$ denotes the decoding order of the sub-message $s_{u,b}$. Meanwhile, $\pi_{u,\hat{v}}$ and $\pi_{v}$ denotes the decoding order of the sub-message $s_{u,\hat{v}}$ and message $s_{v}$.  Particularly, sub-message $s_{u,b}$ will be decoded first, if the decoding order is $\pi_{u,b} < \pi_{u',b'}$ where $\{u' \neq u, b' \neq b\}$ by treating remaining sub-messages/messages as interference. After the decoding of  $s_{u,b}$ is completed, it is removed from the total received signal using SIC, and then the next sub-message/message is decoded according to the adopted decoding order.}
\vspace{-0.3cm}
\subsection{Achievable Rate at CCU due to the Transmission of CEU}
{Each CCU receives the signal from the CEUs during the DT phase. Hence, the achievable rate to decode the received message $s_{v}$ from the CEU-$v$ at CCU-$u$ in a pair $(u,v)$ can be denoted by,}
\begin{equation}
{
    R_{v\rightarrow u}^{[1]}=\delta\log_2\left(1+\frac{|h_{v,u}|^2P_{{v}}}{\hat{I}_{v' \rightarrow u}+\sigma^2}\right),}\label{rate1}
\end{equation}
where{$\hat{I}_{v' \rightarrow u}=\sum_{v' \in \mathcal{V}, v' \neq v}|h_{v',u}|^2P_{v'}s_{v'}$.} 
\subsection{Achievable Rate at the BS}
The BS utilizes decoding order set $\pi_b$ to decode the sub-messages/messages successively after the end of the two transmission phases, specifically, at the end of the CT phase.  {However,  all sub-messages of a particular user do not need to be decoded sequentially. Decoding of the sub-messages is independent of the user and depends on the adopted decoding order. Particularly, if decoding order of stream $s_{u,1}$ is $\pi_{u,1}$ and the decoding order of stream $s_{u',1}$ is $\pi_{u',1}$ and $\pi_{u,1} < \pi_{u',1}$, then $s_{u,1}$ will be decoded before $s_{u',1}$, where $u \neq u'$. The details of the impact of different decoding orders have been provided in Section VI. } Meanwhile, during the CT phase, a total of $\mathcal{U}_s=U \times 3$ sub-messages are received at the BS. After decoding is done, utilizing SIC, the decoded message is removed from the total received signal.  Utilizing Eqn. (\ref{Sig6}), we can find the achievable rate to decode the signal of $s_{u,1}$ at the BS that is received during the CT phase from CCU-$u$ and it can be denoted by,
\begin{equation} \footnotesize
\begin{split}
&{R_{u,1}^{[2]}=(1-\delta)}\\
&\log_2\left(1+\frac{\frac{||\textbf{h}_{u}||^4P_{{u,1}}}{N^2}}{\left(_{(\pi_{u,2} \in \pi_b)}\frac{||\textbf{h}_{u}||^4P_{{u,2}}}{N^2}\right)+\left(_{(\pi_{u,\hat{v}} \in \pi_b)}\frac{||\textbf{h}_{u}||^4P_{{u,\hat{v}}}}{N^2}\right)+\hat{I}_{u',BS}+\frac{|\textbf{h}_{u}^H\boldsymbol{n}_u|^2}{N^2}}\right), \label{rate2}
\end{split}
\end{equation}

\begin{equation} \footnotesize
\begin{split}
R_{u,2}^{[2]}=(1-\delta)
\log_2\left(1+\frac{\frac{||\textbf{h}_{u}||^4P_{{u,2}}}{N^2}}{\left((\pi_{u,\hat{v}} \in \pi_b)\frac{||\textbf{h}_{u}||^4P_{{u,\hat{v}}}}{N^2}\right)+\hat{I}_{u',BS}+\frac{|\textbf{h}_{u}^H\boldsymbol{n}_u|^2}{N^2}}\right),\label{rate2}
\end{split}
\end{equation}
where {$\hat{I}_{u',BS} =\sum_{u' \in \mathcal{U}, u \neq u'}\left(\frac{|\textbf{h}_{u}^H\textbf{h}_{u'}|^2}{N^2}\sum_{b \in \mathcal{B}}{P_{u',b}}+{P_{u',\hat{v'}}}\right)$ and $\pi_{u,1} < \pi_{u,2}, \pi_{u,\hat{v}}$}. 
After successful decoding, and utilizing the SIC, $s_{u,b}$ is removed from the total received signal. During the CT phase, CCU also forwards the signal of its paired CEU to the BS. Hence, the achievable rate to decode the messages of CEU-$v$ due to the transmission of CCU-$u$ can be denoted as,
\begin{equation}\footnotesize
\begin{split}
{
R_{v}^{[2]}=(1-\delta)}
&{\log_2\left(1+\frac{\frac{||\textbf{h}_{u}||^4P_{u,\hat{v}}}{N^2}}{\hat{I}_{u',BS}+\frac{|\textbf{h}_u^H\boldsymbol{n}_u|^2}{N^2}}\right).}
    \end{split}
\end{equation}
{Meanwhile, BS also decodes the message $s_{v}$ that is received during the DT phase from CEU-$v$. Hence, the achievable rate to decode the message, $s_{v}$ can be given by,} 
\begin{equation}
{
R_{v}^{[1]}=\delta\log_2\left(1+\frac{\frac{||\textbf{h}_{v}||^4P_{ {v}}}{N^2}}{\hat{I}_{v',BS}+\frac{|\textbf{h}_{v}^H\boldsymbol{n}_v|^2}{N^2}} \right)}, 
    \label{rate3}
\end{equation}
where {$\hat{I}_{v',BS}=\sum_{v' \in \mathcal{V}, v' \neq v}\frac{|\boldsymbol{h}_{v}^H\boldsymbol{h}_{v'}|^2}{N^2}P_{{{v'}}}$.}\\
Therefore, we can calculate the total achievable rate for the CCU-$u$ as follows, 
\begin{equation}
    R_{u}=\sum_{b \in \mathcal{B}}R_{u,b}^{[2]}.
\end{equation}
After two transmission phases, the total achievable rate for the CEU-$v$ to decode a message, $s_{v}$ can be given by, 
\begin{equation}
{
R_{v}^{tot}=R_{v}^{[1]}+R_{v}^{[2]},}
\end{equation}
However, the total achievable rate of the CEU-$v$ after two transmission phases at the BS should not exceed the achievable rate of the CEU-$v$ at the CCU-$u$ during DT phase, which is denoted as follows,
\begin{equation}
{
R_{v}=\min(R_{v}^{tot}, R_{v\rightarrow u}^{[1]}).}
\end{equation}
\section{Problem Formulation And Solution Roadmap}
\label{Problem Formulation}
\subsection{Problem Formulation}
The main objective is to maximize the sum rate while meeting each UE's QoS requirement in terms of
minimum data rate. Therefore, the sum rate maximization problem with joint pairing, power allocation, and time slot allocation can be formulated as follows:
\allowdisplaybreaks
\begin{subequations}
\label{prob:P1}
\begin{flalign}
\centering
 &\mathcal{P}: \max_{\substack{\boldsymbol{\Psi}, \,\boldsymbol{P}, \,\delta}} \quad \quad \sum_{u \in \mathcal{U}} \sum_{v \in \mathcal{V}}\Psi_{u,v} \left(R_{u}+R_v\right),\:\label{const1} \\
 &\text{s.t.} \quad \Psi_{u,v}\in \{0,1\}, \quad \forall u \in \mathcal{U}, \forall v \in \mathcal{V}, \label{const2}\\
 &\quad \sum_{u \in \mathcal{U}}\Psi_{u,v}= 1, \quad\forall v \in \mathcal{V}, \label{const3}\\
  &\quad \sum_{v \in \mathcal{V}}\Psi_{u,v}= 1, \quad \forall u \in \mathcal{U}, \label{const4}\\
   &\quad R_{u} \ge R_{th,u},\,\, \quad \forall u \in \mathcal{U}, \label{const66}\\
    &\quad \min(R_{v}^{tot}, R_{v\rightarrow u}^{[1]}) \ge R_{th,v},\,\, \quad \forall v \in \mathcal{V}, \forall u \in \mathcal{U}, \label{const6629}\\
    & \quad {P_{u,b}, P_v, P_{u,\hat{v}} \ge 0, \quad \forall u \in \mathcal{U}, \forall v \in \mathcal{V}, \forall b \in \mathcal{B},} \label{const50}\\
    &\quad {\sum_{b \in \mathcal{B}}P_{{u,b}}+P_{{u,\hat{v}}}\le P_{u}^{max},\quad \forall u \in \mathcal{U},}\label{const511}\\
    &\quad {P_{v} \le P_{v}^{max},\quad  \forall v \in \mathcal{V},} \label{const5}\\
     &\quad 0\le \delta\le 1 \label{const7},
\end{flalign}
\end{subequations}
where $\boldsymbol{P}=[P_{u,b}, P_{u,\hat{v}},P_v| \forall u \in \mathcal{U},\forall v \in \mathcal{V}, \forall b \in \mathcal{B}]$ denotes the transmit power of all UEs, and $\boldsymbol{\Psi}$ denotes the user pairing policy, which holds a binary value of 0 or 1. Specifically, $\Psi_{u,v}=1$ implies that CCU-$u$ is paired with CEU-$v$. Meanwhile, $\Psi_{u,v}=0$ represents that CCU-$u$ and CEU-$v$ are not paired. Constraints (\ref{const3}) and (\ref{const4}) ensure that each UE from each group can be paired with only one UE from the other group. Constraints (\ref{const66}) and (\ref{const6629}) ensure that each UE has an achievable rate greater than a minimum achievable rate in order to guarantee the QoS. Constraints (\ref{const511}) and (\ref{const5}) refer to the transmission power budget of CCU-$u$ and CEU-$v$. Finally, (\ref{const7}) represents the time slot duration constraint for the NDF protocol. 
\subsection{Solution Roadmap}
Due to the intractability of problem $\mathcal{P}$, it is very hard to solve it directly. In order to tackle this issue, we adopt a bi-level optimization-based process that divides the original problem into two sub-problems: $\mathcal{P}_{outer}$ and $\mathcal{P}_{inner}.$ In the first sub-problem $P_{outer}$, for given values of transmit powers at the CCUs and CEUs and the time slot duration, we optimize the user pairing policy by developing a low-complexity algorithm that considers semi-orthogonality among the CCUs and chooses the channel with the highest gain between a CCU and a CEU to create the best C-RSMA pairs. In the second sub-problem  $\mathcal{P}_{inner}$, for a given user pair, we optimize the transmit powers of CCUs and CEUs and the time slot duration by designing an SCA-based low-complexity algorithm in order to maximize the sum rate of the whole system. It should be noted that to optimize the time slot allocation, we use an exhaustive search approach. Particularly, we calculate the sum rate of the system for different values of $\delta$, which ranges from 0 to 1. We then choose the value of $\delta$ of those results, which provides the highest sum rate for the system. Finally, the overall problem is solved in three steps: first, for given values of transmit power and time slot duration, we select a CCU for a pair utilizing the SUS algorithm, which will assist in piggybacking the signals of its paired CEU in the CT phase. Next, we select CEU for each pair using a low-complexity MG-based algorithm, considering maximum channel gain between CCU-CEU as a utility function of the matching game. Finally, power allocation of the UEs in a given pair is performed utilizing an SCA-based algorithm in an iterative manner. Specifically, it can be seen that when pairing is done, $\Psi_{u,v}=1$ and $(\boldsymbol{P}^*,\delta^*)$ becomes the optimal solution of the power and time slot duration allocation. Meanwhile, when $\Psi_{u,v}=0$, then $(\boldsymbol{P}^*,\delta^*)$ becomes zero. Thus, for given values of $\boldsymbol{P}^*, \delta^*$, we can write the outer optimization problem as follows:
\allowdisplaybreaks
\begin{subequations}
\label{prob:P3}
\begin{flalign}
\centering
&\mathcal{P}_{outer}: ~ \max_{\substack{\boldsymbol{\Psi}}} \sum_{u \in \mathcal{U}} \sum_{v \in \mathcal{V}}\Psi_{u,v} \left(R_u\left(\boldsymbol{P}^*, \delta^*\right) + R_v\left(\boldsymbol{P}^*, \delta^*\right)\right) ,\: \label{const8} \\
&\text{s.t.} \quad (\ref{const2})-(\ref{const4}).\notag
\end{flalign}
\end{subequations}
Meanwhile, for a given user pairing policy $\boldsymbol{\Psi}^*$, the inner optimization problem can be as follows:
\allowdisplaybreaks
\begin{subequations}
\label{prob:P4}
\begin{flalign}
\centering
 &\mathcal{P}_{inner}: \quad \max_{\substack{\boldsymbol{P},\delta}} \sum_{u \in \mathcal{U}}\sum_{v \in \mathcal{V}}\Psi_{u,v}^*({R}_u(\boldsymbol{P}, \delta )+R_v(\boldsymbol{P}, \delta )),\:  \label{const9}\\
 &\text{s.t.} \quad \quad (\ref{const66})-(\ref{const7}). \label{const11} \notag
\end{flalign}
\end{subequations}
In the following section, we will give the details of the solution approaches of the two sub-problems.
\section{UEs Pairing: Semi-orthogonality and Two-Sided One-to-One Matching Game-based approach} \label{pairing}
Our algorithm for user pairing operates through a two-stage process. Initially, we pick a CCU for each pair using the SUS algorithm. This SUS algorithm considers channel orthogonality among CCUs which assists in reducing IPI. Afterward, we choose a CEU for each pair using an MG-based algorithm that factors in the impact of the channel gains between all CCUs and the corresponding CEU. This channel gain effect serves as a utility function for our selection process and helps to choose CEUs which can maximize the sum rate. The whole process is given below in detail.
\par 1.\textbf{CCU selection}:
We employ a SUS-based algorithm to determine the CCU-$u$ for a pair $(u,v)$. The SUS algorithm is first proposed in \cite{1603708} to design multiple-input multiple-output (MIMO) beamforming. The main idea behind the SUS algorithm is that it tries to choose users with superior channel states and aligned beam directions by using the degree of channel orthogonality among users. The SUS algorithm for CCU selection is provided in Algorithm 1. Our proposed SUS algorithm iteratively selects a subset of size  $\mathcal{U}$ with user channels $\{\boldsymbol{h}_u| u \in\mathcal{U}\}$ that are semi-orthogonal to each other and with relatively large channel gains with the BS. These selected users constitute the group of CCUs. The iteration procedure continues until $U$ number of users are selected or $\mathcal{U}_{0}$ becomes empty. In Algorithm 1, $\mathcal{U}$ represents the set for chosen CCUs, meanwhile, $\mathcal{U}_{0}$ denotes the set of users that are not chosen as CCUs yet.  From the algorithm's steps 3 through 9, for each user $u \in \mathcal{U}_0$ the component of $\boldsymbol{h}_u$ is orthogonal to the subspace covered by $[\boldsymbol{g}_1$, $\boldsymbol{g}_2,\dots,\boldsymbol{g}_{j-1}]$. Afterward, we select the best user with the maximum argument in step 10. It should be noted that in step 12, we specify a SUS factor $\theta$ which ensures that only the users with semi-orthogonal channels remain in the set $\mathcal{U}_{0}$. The value of the SUS factor $\theta$ falls between 0 to 1. Therefore, in step 7, only a small portion of user channel $\boldsymbol{h}_u$ will be projected to the subspace spanned by $[\boldsymbol{g}_1$, $\boldsymbol{g}_2,\dots,\boldsymbol{g}_{j-1}]$. Hence, the users chosen in this way become semi-orthogonal to each other which assists in reducing the IPI, and also relatively large channel gains help them to serve as CCUs. 
\begin{algorithm}[!t]
\DontPrintSemicolon
  \KwInput{$\boldsymbol{h}_k, \forall k \in \mathcal{K}$, SUS factor, $\theta$}
  \KwOutput{Set of CCU, $\mathcal{U}$}
  Initialize $\mathcal{U}= \O$, $\mathcal{U}_{0}= \mathcal{K}$ and $j=1$\;
  \While{$j \le N$ and $\mathcal{U}_{0}\neq \O$}{
  \For{$u \in \mathcal{U}_{0}$}{
  \textbf{if} $(j==1)$ \textbf{then}\;
  \quad $\boldsymbol{g}_u :=\boldsymbol{h}_u$\;
  \textbf{else} \;
  \quad $\boldsymbol{g}_u:=  \boldsymbol{h}_u-\sum_{{j'}=1}^{j-1}\frac{\boldsymbol{h}_j\boldsymbol{g}_{j'}}{\lVert\boldsymbol{g}_{{j'}}\rVert^2}\boldsymbol{g}_{{j'}}$\;
  \textbf{end if}\;
  }
 Select user $j^* := arg \max_{u \in \mathcal{U}_{0}} \lVert\boldsymbol{g}_u\rVert$ \;
  Adjust  $\mathcal{U}\leftarrow \mathcal{U} \bigcup  \{j^*\}$ \;
  Update $\mathcal{U}_{0_{j+1}} \leftarrow \{ u | u \in \mathcal{U}_{0},u \neq j^*, \frac{|\boldsymbol{h}_u \boldsymbol{g}^*_{j}|}{\lVert\boldsymbol{h}_{{u}}\rVert \lVert\boldsymbol{g}_{{j}}\rVert} < \theta\}$ where $\theta$ represents SUS factor which is small positive constant and value varies between $0$ to $1$\;
  $j=j+1$
  }
\caption{SUS-based CCU selection algorithm}
\end{algorithm}
\begin{algorithm}[!t]
\DontPrintSemicolon
 Construct a set of CEU $\mathcal{V}$ with the UEs that are not selected as CCUs, a preference list of CCU, $\mathbb{P}_{u}$, a preference list of CEU, $\mathbb{P}_{v}$, a list of unmatched CCU $\mathbb{U}$, a list of unmatched CEU $\mathbb{V}$\;
  Initialize $\mathbb{U}= \mathcal{U}$, $\mathbb{V}= \mathcal{V}$\;
  \While{$\mathbb{U} \neq 0$ and $\mathbb{V} \neq 0$}{
\For {$i \in \mathbb{V}$}{${v_i} $ proposes itself to the  $u_i \in\mathbb{P}_{v_{i}}$ \;
  \If{$u_i$~ $\mathrm{is~available}$}{
   Match $(u_i,v_i)$ and store the tuple \;
   Remove $u_i$ from $\mathbb{U}$\;
   Remove $v_i$ from $\mathbb{V}$\;}
   \textbf{else}\If{$u_i$ $\mathrm{~is ~already ~paired ~with~} v_i'$ $~\mathrm{(another~ CEU ~and} ~v_i'\neq v_i)$}{
   \If{$\Upsilon(u_i,v_i) > \Upsilon(u_i,v_i')$}{
  unpair $u_i$ from $v_i'$\;
  Match $(u_i,v_i)$ and store the tuple\;
  Remove $u_i$ from $\mathbb{U}$\;
  Add $v_i'$ in $\mathbb{V}$\;
  \Else{
  Keep $v_i$ in $\mathbb{V}$\;
  Keep the pair $(u_i,v_i')$\;}}
  }
  \textbf{else}\If{$(u_i,v_i)$ \rm is a blocking tuple}{
  Match $(u_i,v_i)$ and store the tuple\;
   Remove $u_i$ from $\mathbb{U}$\;
   Remove $v_i$ from $\mathbb{V}$\;
  }
  \Else{
  $v_i$ is rejected by $u_i$\;
  Remove $u_i$ from $\mathbb{P}_{v_{i}}$\;
  }
 }
\caption{SUS-MG-based pairing algorithm}
}
\end{algorithm}
\par 2.\textbf{CEU selection}:
In this subsection, we choose a CEU for each pair by utilizing an MG-based algorithm. We model our CCU-CEU pairing problem as a two-sided one-to-one matching game problem. In fact, the matching game theory is well adapted for a scenario in which two sets of players are paired off in order to produce outcomes that are advantageous to both parties. We set one set of users as proposers and the other sets of users as selectors. Here, we model CEUs as proposers and CCUs as selectors. 
\par \textit{Definition 1: Matching tuple:}
\textit{A one-to-one two-side matching $\Psi$ is a mapping from all the
members of $\mathcal{U}$ into the $\mathcal{V}$ satisfying the following conditions such that
\begin{enumerate}[label=(\alph*)]
\item $\Psi(v) \in \mathcal{U}, \Psi(u) \in \mathcal{V}$,
\item $\Psi(u) = v \Leftrightarrow \Psi(v) = u, \forall u \in \mathcal{U}, \forall v \in \mathcal{V}$, 
\item $|\Psi(v)| = 1, |\Psi(u)| = 1, \forall u \in \mathcal{U}, \forall v \in \mathcal{V}$.
\end{enumerate}
}
\par Condition (a) indicates that the matching partner of one set is a member of another set, (b) indicates that if $u$ matches with $v$ then $v$ matches with $u$ as well, and finally, (c) suggests that each CCU can match with only one CEU and vice versa.
\par \textit{Definition 2: Preference utility: In a matching game, the design of preference utility assists in finding the best possible match, which can maximize the objective function.
}
\par We design the utility function of our proposed matching game based on the channel gains between CEU-$v$ and all CCUs in $\mathcal{U}$. We denote the preference utility of any tuple of the CCU-CEU pair
$(u,v)$ by considering the channel gain between the CEU-$v$ and all the CCUs in order to be considered to be a potential matched tuple. Hence, we can present the preference utility of a matching tuple $(u,v)$ as follows:
\begin{equation}
    \Upsilon(u,v)=arg \max \{{|h_{v,u}|}\}, \forall u \in \mathcal{U}, \forall v \in \mathcal{V}.
\end{equation}
The main idea behind choosing such
utility function comes from the objective of $\mathcal{P}$,
where each matching tuple constructs a CCU-CEU pair such that the sum-rate of the overall system is maximized. Since every CCU and CEU sustain distinctive channel gains with the BS and the channel gain between each CCU and CEU is unique, the utility function results in a unique utility value for every pair of CCU-CEU. 
In every matching theory, an important factor is the change of the matching pair over time. This is because there exists a competition among CEUs to be paired with an individual CCU.  Specifically, if there exists a matching tuple CCU-CEU $(u,v)$ and CCU-$u$ receives a proposal from CEU-$v'$ for pairing, CCU-$u$ chooses CEU-$v'$ over CEU-$v$ if and only if the preference utility $\Upsilon(u,v') > \Upsilon(u,v)$. In this situation, CCU-$u$ rejects CEU-$v$ and creates a tuple with CEU-$v'$. 
\par \textit{Definition 3: Preference List:
Each participant can create his own descending-ordered preference profile by assessing the utilities of the various tuples. Using this preference profile, CEU can identify its
preference from the set of CCUs.}  
\par Let $\mathbb{P}_v=[u_1,u_2,\dots,u_U]$ is a preference list of CEU-$v$, where $u_1$ is considered as the most preferred user to be paired with CEU-$v$. On the other hand, $u_U$ represents the least preferred user to be paired with CEU-$v$. In our proposed scheme, each CEU-$v$ prepares a preference list of its preferred CCU based on that channel gain between CEU-$v$ and all CCUs. If CEU-$v$ sustains the highest channel gain with CCU-$u$, then CCU-$u$ is put into the first place of CEU-$v$'s preference profile.
\par \textit{Definition 4: Blocking pair:
 The pair $(u,v) \in (\mathcal{U} \times \mathcal{V})$ is said to block in a  matching $\Psi$ if
 the utility of $(u,v)$ is higher than all other possible pairings according to $v$'s preference profile.} 
\par For example, $u$ is listed as a preferred CCU at preference profile $\mathbb{P}_v$ of CEU-$v$. At the same time $u$ is also listed as a preferred CCU at the preference profile $\mathbb{P}_{v'}$ of CEU-$v'$. In addition, CCU-$u$ prefers CEU-$v$ over CEU-$v'$ because it sustains the highest preference utility of all possible matching pairs. In this case, $(u,v)$ blocks the matching of $(u,v')$ and $(u,v)$ should be matched together in all situations.
\par \textit{Definition 5: Stable matching: A matching $\Psi$ is defined as pairwise stable
if it is not blocked by any blocking pair.}
\par 
In our proposed scheme, we aim to seek stable matching, and the concept
of stable matching is as follows: each CEU tries to match with its most preferable CCU and the CCU tends to choose the CEU that can maximize the total utility. First, each CEU proposes itself to its most favorable CCUs. Each CCU then receives offers from the CEU. Based on CCU's own preference, it can accept or reject the offer. It should be noteworthy that each CCU may receive offers from multiple CEUs. In order to remove the conflict between the users, we invoke the concept of a blocking pair. Specifically, if a blocking pair exists in the system, CCU will stay paired with that particular CEU regardless of whatever proposals are coming over. However, if there does not exist any blocking pair, then matched tuple $(u,v)$ is considered stable.
The detailed SUS-MG-based pairing policy is provided in Algorithm 2. 
\vspace{-0.3cm}
\section{Power and Time Slot Duration Allocation for Each C-RSMA Pair} \label{SCA}
In this section, our objective is to maximize the sum-rate at all UEs for given values of time slot duration $\delta$. We solve the power allocation problem utilizing the SCA-based approach.
\par $\mathcal{P}_{inner}$ is a non-convex optimization problem due to the existence of objective at \eqref{const9} and constraints \eqref{const66}, and \eqref{const6629}. To handle the non-convexity in the objective, we introduce an auxiliary variable $\boldsymbol{\Lambda}=[\Lambda_u|\forall u \in \mathcal{U}, \Lambda_v | \forall v \in \mathcal{V}]$ and the objective and constraints \eqref{const66}, and \eqref{const6629} can be written as follows,
\begin{equation}
\hat{\mathcal{P}}_{inner}: \max_{\boldsymbol{\Lambda},\boldsymbol{P},\delta} \sum_{u \in \mathcal{U}}\sum_{v \in \mathcal{V}}(\Lambda_u+\Lambda_v), \label{eq1}
\end{equation}
\begin{equation}
\sum_{b \in \mathcal{B}}(1-\delta)\log_2 (1+\alpha_{u,b}) \ge \Lambda_u, \label{eq2}
\end{equation}
\begin{equation}
    \Lambda_{u} \ge R_{th,u}, \label{eq3}
\end{equation}
\begin{equation} 
{
    \delta\log_2 (1+\beta_{v})+
(1-\delta)\log_2 (1+\beta_{u}) \ge \Lambda_v,} \label{eq4}
\end{equation}
\begin{equation}
{
    \delta \log_2(1+\omega_{v}) \ge \Lambda_v,} \label{eq5}
\end{equation}
\begin{equation}
{
    \Lambda_v \ge R_{th,v},}\label{eq6}
\end{equation}
\begin{equation} \footnotesize
\frac{\frac{||\textbf{h}_{u}||^4P_{{u,b}}}{N^2}}{{\frac{||\textbf{h}_{u}||^4P_{{u,b'}}}{N^2}}+\left(\frac{||\textbf{h}_{u}||^4P_{{u,\hat{v}}}}{N^2}\right)+\hat{I}_{u',BS}+\frac{|\textbf{h}_{u}^H\boldsymbol{n}_u|^2}{N^2}} \ge \alpha_{u,b}, \label{approx2}
\end{equation}
\begin{equation}
\frac{\frac{||\textbf{h}_{v}||^4P_{ {v}}}{N^2}}{ \hat{I}_{v',BS}+\frac{|\textbf{h}_{v}^H\boldsymbol{n}_v|^2}{N^2}}  \ge \beta_{v}, \label{SCA2}
\end{equation}
\begin{equation} \frac{\frac{||\textbf{h}_{u}||^4P_{{u,\hat{v}}}}{N^2}}{\hat{I}_{u',BS}+\frac{|\textbf{h}_{u}^H\boldsymbol{n}_u|^2}{N^2}} \ge \beta_{u},\label{SCA3}
\end{equation}
\begin{equation}
    \frac{|h_{v,u}|^2P_{{v}}}{\hat{I}_{v' \rightarrow u}+\sigma^2}  \ge \omega_v \label{SCA31}
\end{equation}
where $\boldsymbol{\alpha}=[\alpha_{u,b}| \forall u \in \mathcal{U}, \forall b \in \mathcal{B}]$, $\boldsymbol{\beta}=[\beta_{u}, \beta_{v}| \forall u \in \mathcal{U}, \forall v \in \mathcal{V}, \forall b \in \mathcal{B}]$ and $\boldsymbol{\omega}=[\omega_v| \forall v \in \mathcal{V}]$.
However, \eqref{approx2} is still non-convex. Hence, we introduce slack variables $\boldsymbol{\gamma}=[\gamma_{u,b}| \forall u\in \mathcal{U}, \forall b \in \mathcal{B}]$ and replace the interference in \eqref{approx2}  with this slack variable. We can rewrite \eqref{approx2} as follows,
 \begin{equation}
 {
   \frac{1}{N^2}\left(\frac{||\textbf{h}_u||^4P_{u,b}}{\gamma_{u,b}}\right) \ge \alpha_{u,b},}\label{approx4}
\end{equation}
\begin{equation}
{\gamma_{u,b}\ge{\frac{||\textbf{h}_{u}||^4P_{{u,b'}}}{N^2}}+\left(\frac{||\textbf{h}_{u}||^4P_{{u,\hat{v}}}}{N^2}\right)+\hat{I}_{u',BS}+\frac{|\textbf{h}_{u}^H\boldsymbol{n}_u|^2}{N^2},}\label{approx5}
\end{equation}
Since \eqref{approx4} is still non-convex, according to arithmetic and geometric means (AGM) inequality \cite{7946258} for any non-negative variables $x,y$ and $z$, and if $xy \le z$ then the $2xy \le (ax)^2+(\frac{y}{a})^2 \le 2z$, where the first inequality holds if and only if $a=\sqrt{y/x}$. Based on this, equations 
\begin{equation}\footnotesize
{
\frac{1}{N^2}* ||\textbf{h}_u||^4P_{u,b} \ge \alpha_{u,b}\gamma_{u,b},}
\label{approx6}
\end{equation}
\begin{equation}
{
    \frac{1}{N^2}*2 ||\textbf{h}_u||^4P_{u,b} \ge (\alpha_{u,b}*\phi_{u,b})^2+(\phi_{u,b}/\gamma_{u,b})^2,} \label{approx7}
\end{equation}
{where $\phi_{u,b}=\sqrt{\gamma_{u,b}/\alpha_{u,b}}$ and $\phi_{u,b}$ should be updated iteratively.}
\eqref{SCA2}, \eqref{SCA3}, and \eqref{SCA31} can be handled similarly as \eqref{approx6} by AGM inequality. Based on the above discussions and approximations, we can rewrite $\hat{\mathcal{P}}_{inner}$ as follows, 
\allowdisplaybreaks
\begin{subequations} 
\label{prob:Pf}
\begin{align}
\centering
& \hat{\mathcal{P}}_{inner}: \max_{\boldsymbol{\Lambda},\boldsymbol{P},\delta, \boldsymbol{\alpha}, \boldsymbol{\gamma}, \boldsymbol{\beta}, \boldsymbol{\eta}, \boldsymbol{\mu}} \sum_{u \in \mathcal{U}}\sum_{v \in \mathcal{V}}(\Lambda_u+\Lambda_v), \\
 &{\text{s.t.} \quad c_1: \frac{1}{N^2}* 2*||\textbf{h}_v||^4P_{v} \ge (\beta_{v}*\phi_v)^2+(\phi_v/\mu_{v})^2,} \\
 & \quad {c_2: \mu_v \ge \hat{I}_{v',BS}+\frac{|\textbf{h}_{v}^H\boldsymbol{n}_v|^2}{N^2},}\\
 & \quad {c_3: \frac{1}{N^2}* 2*||\textbf{h}_u||^4P_{u,\hat{v}} \ge (\beta_{u}*\phi_u)^2+(\phi_u/\mu_{u})^2,}\\
& \quad {c_4: \mu_u \ge \hat{I}_{u',BS}+\frac{|\textbf{h}_{u}^H\boldsymbol{n}_u|^2}{N^2},}\\
 & \quad {c_4:2*|h_{v,u}|^2*P_{v} \ge (\omega_v*\phi_v)^2+(\phi_v/\eta_{vu})^2}\label{approx1011},\\
 & \quad {c_5:
    \eta_{vu} \ge \hat{I}_{v'\rightarrow u}+\sigma^2,} \\
& \quad {\eqref{eq1}-\eqref{eq6}, \eqref{approx5}, \eqref{approx7}.}\notag
\end{align}
\end{subequations}
where $\boldsymbol{\mu}=[\mu_{v}, \mu_{u}|\forall v \in \mathcal{V}, \forall u \in \mathcal{U}]$, $\boldsymbol{\zeta}=[\zeta_{v}| \forall v \in \mathcal{V}]$, $\boldsymbol{\eta}=[\eta_{vu}| \forall v \in \mathcal{V}]$. 
The problem denoted as $\mathcal{\hat{P}}_{inner}$ is a convex second-order cone program (SOCP), which can be efficiently addressed using various convex optimization solvers like YALIMP or CVX. Based on the above analysis, the proposed SCA-based algorithm $\mathcal{\hat{P}}_{inner}$  is provided in Alg. 3. The overall algorithm for the proposed system is provided in Alg. 4. 
\begin{algorithm}[!t]
\caption{Proposed SCA-based power allocation algorithm}\label{alg:3}
\textbf{Input}: Time slot duration $\delta$, tolerance $\epsilon$, pairing variable $\boldsymbol{\Psi}$\\
Initialize initial feasible points $\boldsymbol{P}^0$, $\boldsymbol{\Lambda}^0$, $\boldsymbol{\gamma}^0$, $\boldsymbol{\zeta}^0$, $\boldsymbol{\mu}^0$, $\boldsymbol{\eta}^0$, 
$j=0$;\\
\For {$\delta=0.1:0.1: 1$}{
\While{$|\boldsymbol{\Lambda}^j - \boldsymbol{\Lambda}^{j-1}| < \epsilon$ }{
  $j=j+1$ ;\\
  solve $\hat{\mathcal{P}}_{inner}$ using  $\boldsymbol{P}^{j-1},\boldsymbol{\Lambda}^{j-1},\boldsymbol{\zeta}^{j-1}, \boldsymbol{\gamma}^{j-1}, \boldsymbol{\mu}^{j-1}, \boldsymbol{\eta}^{j-1}$ and denote optimal objective as $\boldsymbol{\Lambda}^*$ and the optimal variables as $\boldsymbol{P}^*, \boldsymbol{\Lambda}^*, \boldsymbol{\gamma}^{*},\boldsymbol{\zeta}^*, \boldsymbol{\mu}^*, \boldsymbol{\alpha}^*, \boldsymbol{\eta}^*$ ; \\
  update $\boldsymbol{P}^j$ $\leftarrow \boldsymbol{P}^*$,  $\boldsymbol{\Lambda}^j$ $\leftarrow \boldsymbol{\Lambda}^*$,  $\boldsymbol{\zeta}^j$ $\leftarrow \boldsymbol{\zeta}^*$, $\boldsymbol{\mu}^j \leftarrow \boldsymbol{\mu}^*$, $\boldsymbol{\eta}^j \leftarrow \boldsymbol{\eta}^*$, $\boldsymbol{\gamma}^j \leftarrow \boldsymbol{\gamma}^*$ ;
}
$R_{\delta}= \boldsymbol{\Lambda}^*$;
}
$R_{opt}=\max(R_{\delta})$;
\end{algorithm}
\vspace{-0.3cm}
\subsection{Computational complexity analysis}
In order to measure the computational complexity of Algorithm 4, we need to analyze the complexity of the
$\mathcal{P}_{outer}$ and $\mathcal{P}_{inner}$. $\mathcal{P}_{outer}$ is solved using the SUS-MG algorithm. Hence, it depends on the complexity of SUS and MG individually. Meanwhile, $\mathcal{P}_{inner}$ is solved utilizing the SCA-based method in every step of the exhaustive search to find an optimized $\delta$. Hence, the complexity of $\mathcal{P}_{inner}$ depends on the step size of the exhaustive search and proposed SCA-based approach.
The SUS algorithm iterates over all $K$
user channels, and chooses the channel orthogonal to the sub-space spanned by the already selected user channels in an iterative manner. Hence, the computational complexity of the SUS algorithm is $\mathcal{O}(K^2)$. However, the orthogonality threshold $\theta$ accelerates the convergence by reducing the search space which results in a much lower complexity in practice. The complexity of the proposed matching-based algorithms depends on CEU's preference profile creation and CEU-CCU's proposing and selecting process. Specifically, each CEU proposes itself to the CCU based on its preference profile. The CEU can propose itself to its preferred CCU and the CCU can accept or reject the proposal based on the preference utility. The sorting for creating preference profiles is based on quick-sort and its complexity is $\mathcal{O}(n\log (n))$. For the proposing-selecting process, there is no more than $U$ number of CCUs, and one CEU-$v$ in each cell can perform the proposing process with $U$ number of CCUs. Therefore, the maximum number proposing-selecting operation is $VU$. Let us assume that $N_{it}$ represents the total number of iterations if there exists no blocking pair.  Hence, the total complexity of the SUS-MG algorithm can be calculated as $\mathcal{O}(K^2+N_{it}VU)$.\\
It is important to highlight that we perform an exhaustive search within the range of values between 0 and 1, with a step size of 0.1. During each step of this exhaustive search, we utilize Algorithm 3 to find a solution. Our SCA-based algorithm is a SOCP that has the complexity of $(S_1^2S_2)$, where $S_1 = (9 + N)K$ is the total number of variables and $S_2 = 14K$ is the total number of
constraints. Thus, the total complexity of Algorithm 3 is $\mathcal{O}(JN_t^2K^{3.5}log_2(1/\epsilon))$ where $J$ represents the total number of steps for exhaustive search. Hence, the total complexity of the overall algorithm is $\mathcal{O}(K^2+N_{it}VU+JN^2K^{3.5}log_2(1/\epsilon))$.
\begin{algorithm}[!t]
\DontPrintSemicolon
Select CCU using Algorithm 1. \;
Calculate preference utility for all combinations of CCU-CEU pair.\;
Construct CCU-CEU pair using Algorithm 2.\;
Solve  power allocation problem with Algorithm 3.\;
\caption{Overall algorithm}
\end{algorithm}
\section{Simulation Results \& Discussions}
\label{Simulation}
In this section, extensive simulations are carried out to evaluate the performance of the proposed uplink C-RSMA MISO system. The simulation parameters are summarized in Table II. The channel model includes small-scale fading and path loss. Particularly, the small-scale fading follows the Rayleigh distribution with unit variance. {Unless otherwise specified, we assume that channel gains $\boldsymbol{h}_u, \boldsymbol{h}_v, \boldsymbol{h}_{v,u}$ follow the exponential distribution $\lambda_u, \lambda_v, \lambda_{v,u}$. The values of  $\lambda_u, \lambda_v, \lambda_{v,u}$ are 15 dB, 7dB, and 12 dB respectively.} Moreover, we assume that the path loss factor is calculated in terms of channel disparities between the BS and UEs such that $\tau=0.1$ represents the high channel disparity and $\tau=1$ represents a low channel disparity \cite{9771468, 9123680}. To calculate the channel disparities of UEs,  we decrease the channel disparity uniformly from 1 with step size  $\frac{1}{K}$. For example, when $K=6$, $\tau_1=1$, $\tau_2 = 0.83$, $\tau_3 = 0.66$, $\tau_4 = 0.49$, $\tau_5 = 0.32$, and $\tau_6 = 0.15$.  For the sake of comparison, we compare the following strategies with our proposed system. 
\begin{itemize}
    \item{\textit{C-RSMA Random}: In this approach, both the CCU and CEU selection processes is random. The cooperation is performed between CCU and CEU in NDF HD mode. It should be noted that the power allocation is performed using the proposed SCA approach.}
    \item {\textit{C-NOMA fixed $\delta$=0.5 SUS-MG}: In this approach, CCU and CEU are selected using the proposed approach, and cooperation is performed between CCU and CEU in decode-and-forward (DF) HD mode. The power allocation is performed using the proposed SCA approach. However, in this approach, we adopted C-NOMA instead of C-RSMA as the MA scheme \cite{10225434}}.
    \item {\textit{RSMA SUS-MG}: In this approach, we utilize our proposed SUS-MG strategy to create a CCU-CEU pair and power allocation to UEs is performed using SCA. However, there is no cooperation takes place between UEs and we adopt the general RSMA scheme \cite{9257190}.}
    \item {\textit{NOMA SUS-MG}: In this approach, a general uplink NOMA strategy without cooperation is adopted where we utilize our proposed SUS-MG and SCA-based scheme to create a CCU-CEU pair and power allocation to UEs \cite{8876877}.}
\end{itemize}
\begin{table}
\centering
\caption{Simulation parameters}
\begin{tabular}{ |l|c|c| }
 \hline
  \textbf{Parameter} & \textbf{Symbol} & \textbf{Value} \\
  \hline
Number of antennas at BS & $N$ & 8 \\
 \hline
Number of users & $K$ & 6 \\
 \hline
 Rate threshold of CCUs & $R_{th,u}$ & {0.5 bps/Hz} \\
 \hline
Rate threshold of CEUs & $R_{th,v}$& {0.1 bps/Hz} \\
 \hline
 SUS factor & $\theta$ & 0.4\\
 \hline
\end{tabular}
\end{table}
\subsection{{Convergence of proposed scheme}}
\begin{figure}[!t]
    \centering
\includegraphics[width=0.8\columnwidth]{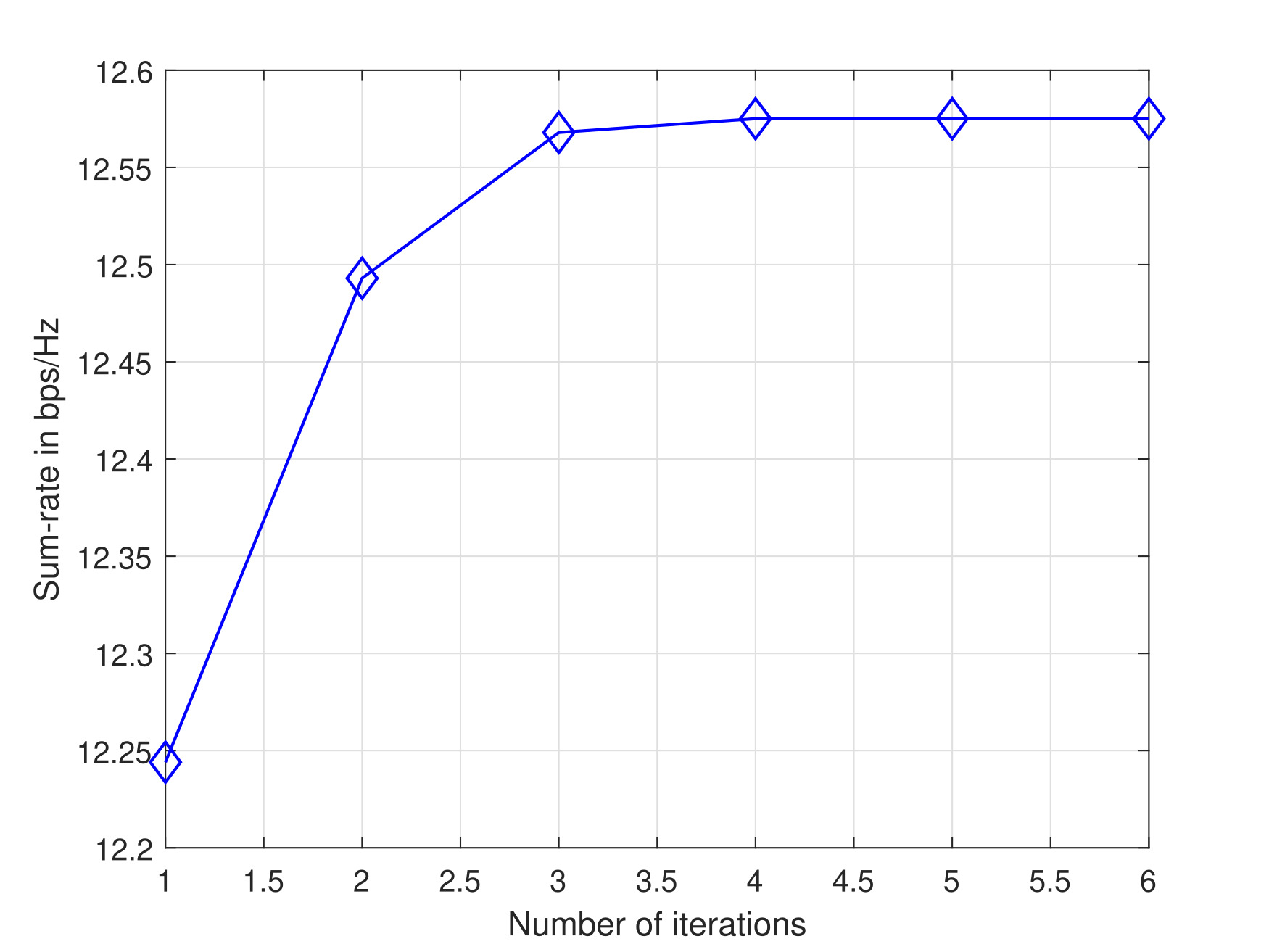}
\caption{{Sum rate vs. number of iterations}}
\label{convergence}
\end{figure}

{Fig. \ref{convergence} shows the convergence behavior of our proposed algorithm with the system parameters: number of UEs $K=6$, $R_{th,1}$ = 0.5 bps / Hz, and $R_{th,2}$ = 0.1 bps / Hz, number of antennas at BS $N=8$, a power budget of the CCUs is 23 dBm and the power budget of the CEUs is 20 dBm. We have plotted the graph with the sum rate (our objective) versus the number of iterations it takes to converge the proposed algorithm. It can be observed that the proposed
uplink HD C-RSMA algorithm converges within around 4–5 iterations.}
\subsection{{Impact of message splitting on UEs}}
{We investigate the impact of message splitting on CCUs and CEUs to investigate the advantage of C-RSMA in uplink. We performed an experiment where we took four scenarios of C-RSMA and two scenarios of RSMA to choose the best possible case for splitting or not splitting of messages.  Details of the investigated splitting process are provided below:}
\begin{figure}[!t]
    \centering
\includegraphics[width=0.85\columnwidth]{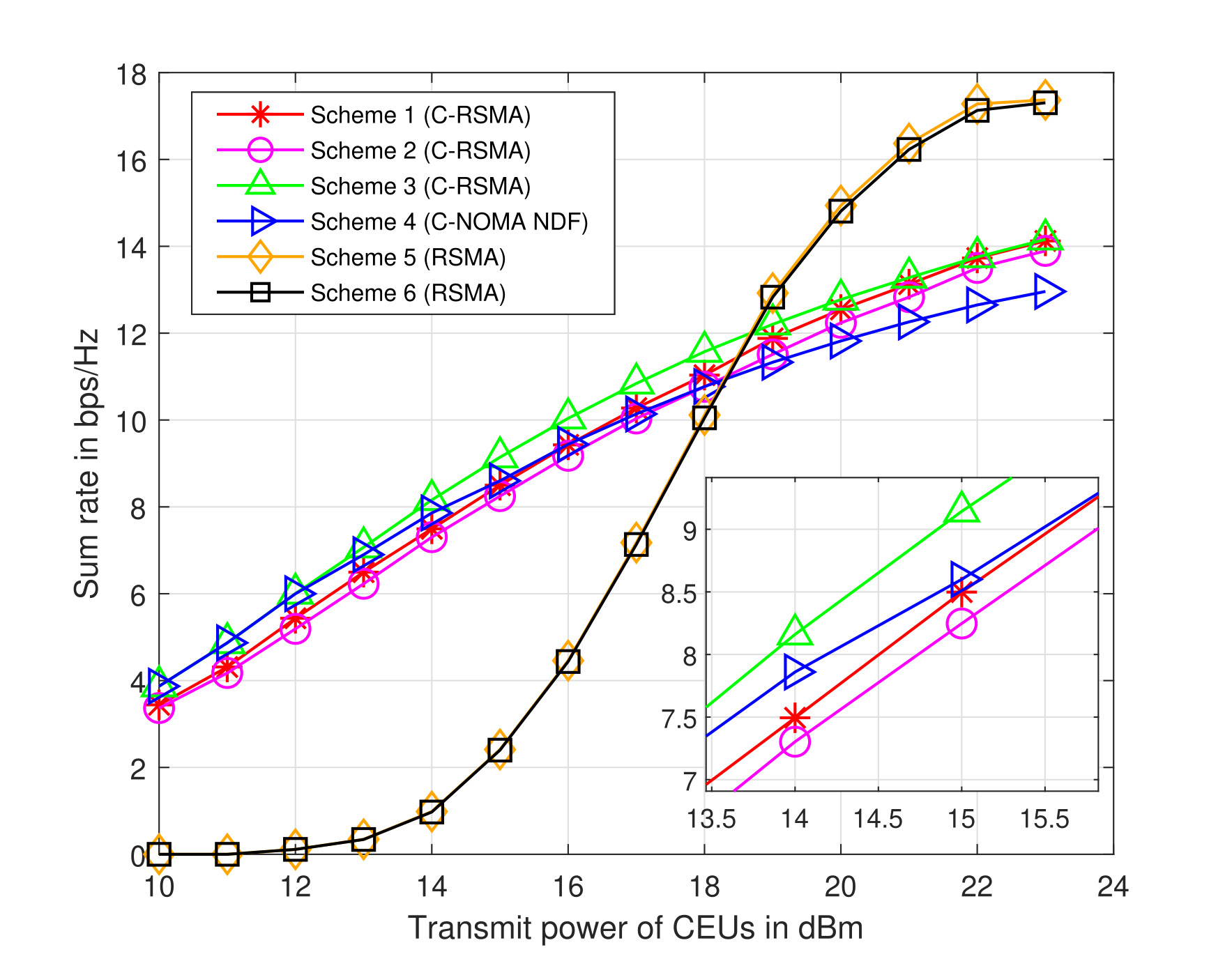}
\caption{{Sum rate vs. power budget of CEUs with decoding order 3}}
\label{split_compare_decode_21}
\end{figure}
\begin{figure}[!t]
    \centering
\includegraphics[width=0.85\columnwidth]{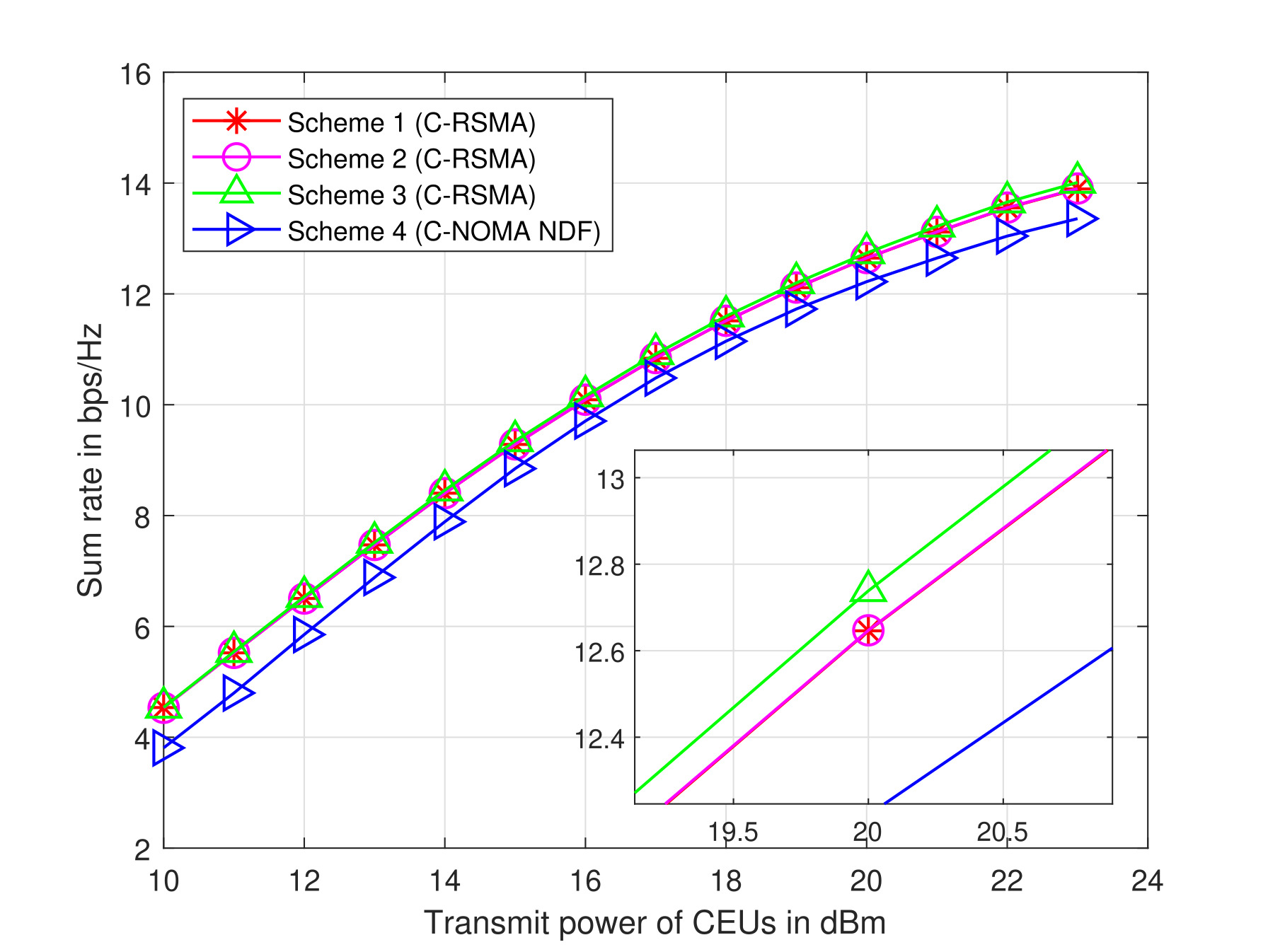}
\caption{{Sum rate vs. power budget of CEUs with decoding order 1}}
\label{split_compare_decode_11}
\end{figure}
\begin{itemize}
    \item {{\textbf{Scheme 1, C-RSMA $2K-1$ split:}
    In this scheme, we split all CCU messages into two sub-messages and all CEU messages into two sub-messages except one CEU message. Particularly, one CEU's message is kept without splitting.} 
    \item \textbf{Scheme 2, C-RSMA $2K$ split:} In this scheme, we split all CCU and CEU messages into two sub-messages. 
    \item \textbf{Scheme 3, C-RSMA no split on CEUs:} In this scheme, all CCU messages are split into two sub-messages. Meanwhile, all CEU messages are kept without splitting.
    \item \textbf{Scheme 4, No split on CCUs and CEUs (C-NOMA NDF))}: In this scheme, we do not split the messages of CCUs and CEUs at all. This scheme is equivalent to C-NOMA with NDF protocol.
    \item \textbf{Scheme 5, RSMA $2K$ split:} In this scheme, we consider a general RSMA scheme with no cooperation. All CCU and CEU messages have been split into two sub-messages.
     \item \textbf{Scheme 6, RSMA $2K-1$ split:} In this scheme, we consider a general RSMA scheme with no cooperation. All CCU messages have been split into two sub-messages. Meanwhile, all CEU messages except one CEU have been split into two sub-messages.
    }
\end{itemize}
{For Fig. \ref{split_compare_decode_21}, in terms of decoding order at the BS, we followed decoding order 3 (details of decoding order presented in the following sub-section).
Fig. \ref{split_compare_decode_21} shows the performance comparison among the four above-mentioned schemes for C-RSMA and two schemes of RSMA while varying the power budget of CEUs. It can be seen that when the power budget of CEUs is low, Scheme 1 and Scheme 2 achieve a lower sum rate  than Scheme 3 and Scheme 4. This is because when the power budget of CEUs is low, Scheme 1 and Scheme 2 do not benefit from splitting user messages into multiple parts by leveraging the flexible interference management process during the decoding. More particularly, when the power budget of CEUs is low, the sub-messages fail to transmit any useful information by overcoming the interference in the DT phase. However, when the power budget of CEUs becomes high, splitting user messages gives more benefits during the DT phase as it can overcome intra- and inter-pair interference and can leverage the benefits of a flexible decoding process at the BS. On the other hand, Scheme 3 and Scheme 4, when the power budget is low, achieve almost similar performance. However, when we increase the power budget of CEUs, the rate of Scheme 4 tends to drop, and at 20 dBm, it drops below all other schemes. This is because as no message is split in Scheme 4, during the decoding process, the messages are decoded as a whole. Hence, it fails to leverage the benefits of flexible decoding as C-RSMA. On the other hand, all three other schemes can benefit from the flexible decoding process due to the splitting of messages. Particularly, when a user message is split into two sub-messages/streams and the sub-messages need not be decoded sequentially; hence, one sub-message is decoded with higher interference, and another one can be decoded with lower interference, which results in a higher rate.}{On the other hand, one can see that the sum rate of general RSMA with $2K$ (Scheme 5) and $2K-1$ split (Scheme 6) overlaps with each other. It proves that $2K-1$ splitting of messages can achieve the capacity region. Hence, $2K$ splitting for RSMA is unnecessary. However, from our observation from the above experiment, this phenomenon does not hold for C-RSMA.} {Moving on to Fig. \ref{split_compare_decode_11}, we adopted decoding order 1 (details of decoding order 1 is given in the following sub-section) to evaluate the performance of four C-RSMA schemes. It can be seen from the figure that three C-RSMA schemes achieve almost similar performance due to the flexible decoding process of RSMA. However, Scheme 4 achieves lower performance than all other schemes. This is because decoding order 1 adopts a decoding process where CEUs messages are decoded first. Hence, the decoding of CEU messages suffers from higher interference from the CCUs, resulting in a lower sum rate. On the other hand, even though CEU messages are not split in Scheme 3, the split of CCU messages helps Scheme 3 achieve better performance utilizing decoding order 1. From our above experiments and observations, we can conclude that in uplink C-RSMA, it is better to split only one user message in each pair and keep other user messages without splitting to get better benefit from the splitting and flexible interference management process.  } 

\subsection{Impact of decoding orders in uplink C-RSMA}
We heuristically investigate the decoding order of the sub-messages/messages at the BS for the uplink C-RSMA framework. We investigate three decoding orders of sub-messages in order to choose the decoding order of sub-messages that provides the higher sum rate. In the following description, we denote a pair as $uv_l$ and their index number as $[1,2,\dots, L]$, $L$ is the total number of pairs. 
Details of the investigated three decoding orders are provided below:
\begin{itemize}
    \item {\textbf{Decoding order 1}: The decoding order followed at BS to decode the sub-messages/messages of two phases for $K=6$ users are given as:  $\pi_{BS}=[\pi_{u,\hat{v}} (uv_1), \pi_{v} (uv_1)  <\pi_{u,\hat{v}} (uv_2), \pi_{v} (uv_2)  < \pi_{u,\hat{v}}(uv_3),\pi_{v} (uv_3)  < \pi_{u,1}(uv_1)<\pi_{u,1}(uv_2)<\pi_{u,1}(uv_3)<\pi_{u,2}(uv_1) <\pi_{u,2}(uv_2)<\pi_{u,2}(uv_3)]$. The above decoding order suggests that at BS, the message of CEU of pair $uv_1$, is decoded first. Since $s_{u,\hat{v}}$ and $s_{v}$ correspond to the same message, they will be decoded together. Then, it is removed from the total received signal using SIC. In this way, we decode all the messages of all CEUs of all pairs. Then, the original sub-messages of all CCUs of the CT phase are decoded. It should be noted that each time one sub-message/message is decoded, it is removed utilizing SIC from the total received signal. Hence, the next sub-message/message to be decoded encounters less interference.
\item \textbf{Decoding order 2}: For this decoding order, we followed the following decoding order at BS: $\pi_{BS}=[\pi_{u,1} (uv_1) < \pi_{u,2} (uv_1) < \pi_{u,\hat{v}} (uv_1), \pi_{v} (uv_1)<\pi_{u,1} (uv_2) < \pi_{u,2} (uv_2) < \pi_{u,\hat{v}} (uv_2),  \pi_{v} (uv_2) < \pi_{u,1} (uv_3) < \pi_{u,2} (uv_3) < \pi_{u,\hat{v}} (uv_3),\pi_{v} (uv_3)]$. Specifically, in this decoding order, we decode the sub-messages pairwise sequentially. First, we decode all sub-messages of CCU-$u$ of pair $uv_1$. Then, we decode the message of CEU-$v$ of pair $uv_1$. Then, we decode the sub-messages of the next pair, and so on.  Each time we decode one sub-message, it is removed using SIC.
\item \textbf{Decoding order 3}: For this decoding order, we adopted the following decoding order: $\pi_{BS} = [\pi_{u,1} (uv_1) < \pi_{u,1} (uv_2) < \pi_{u,1} (uv_3) < \pi_{u,2} (uv_1)< \pi_{u,2} (uv_2) < \pi_{u,2} (uv_3) < \pi_{u,\hat{v}} (uv_1), \pi_{v} (uv_1) < \pi_{u,\hat{v}} (uv_2), \pi_{v} (uv_2)< \pi_{u,\hat{v}} (uv_3),\pi_{v} (uv_3) ]$. Each time we decode one sub-message, it is removed utilizing SIC.}
\end{itemize}
\begin{figure}[!t]
\centering
\includegraphics[width=0.8\columnwidth]{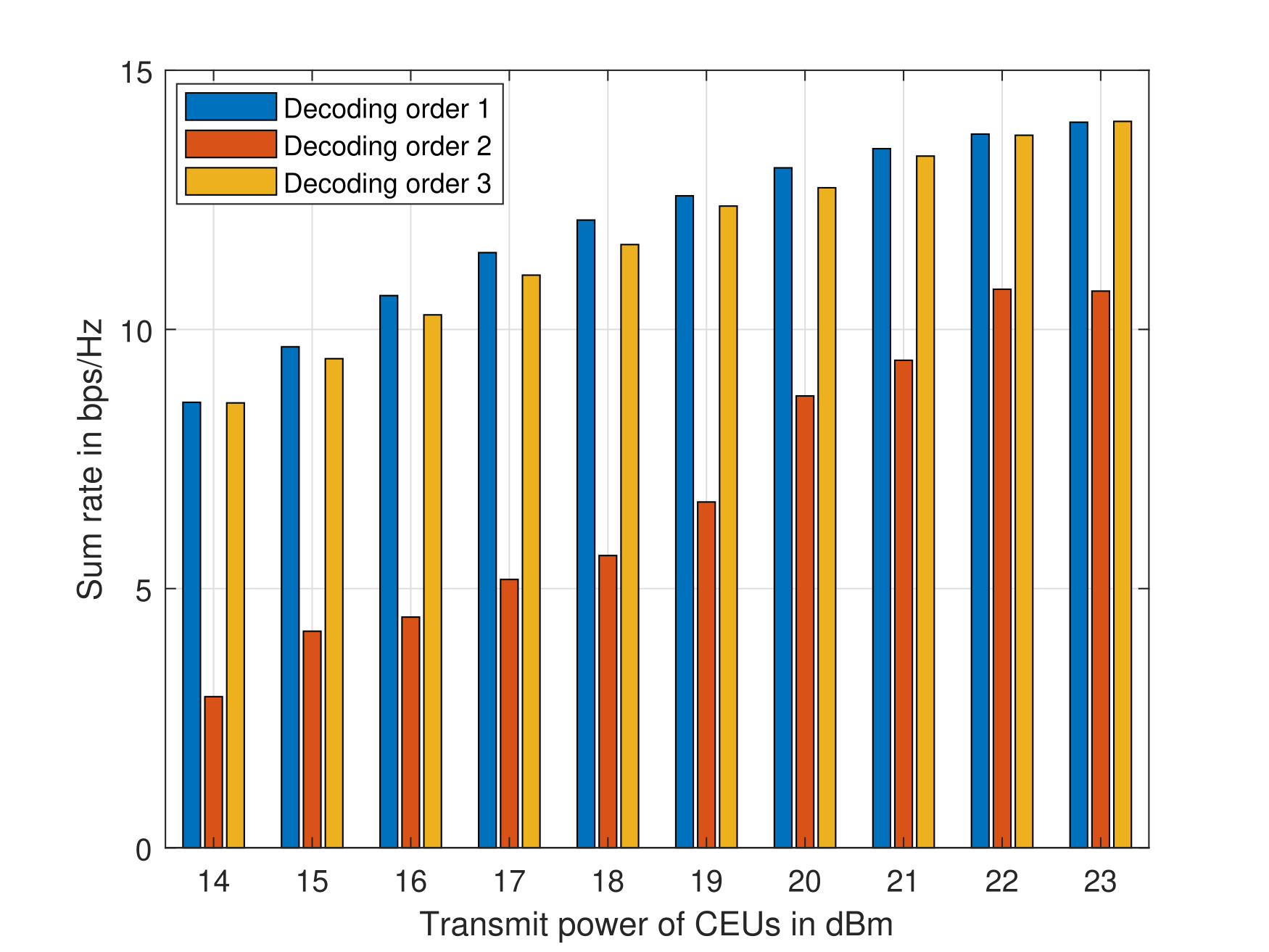}
    \caption {{Sum rate vs different decoding orders while varying power budget of CEUs and $P_u^{max}=23$ dBm}}
    \label{decodingorder21}
\end{figure}
\begin{figure}[!t]
    \centering 
    \includegraphics[width=0.8\columnwidth]{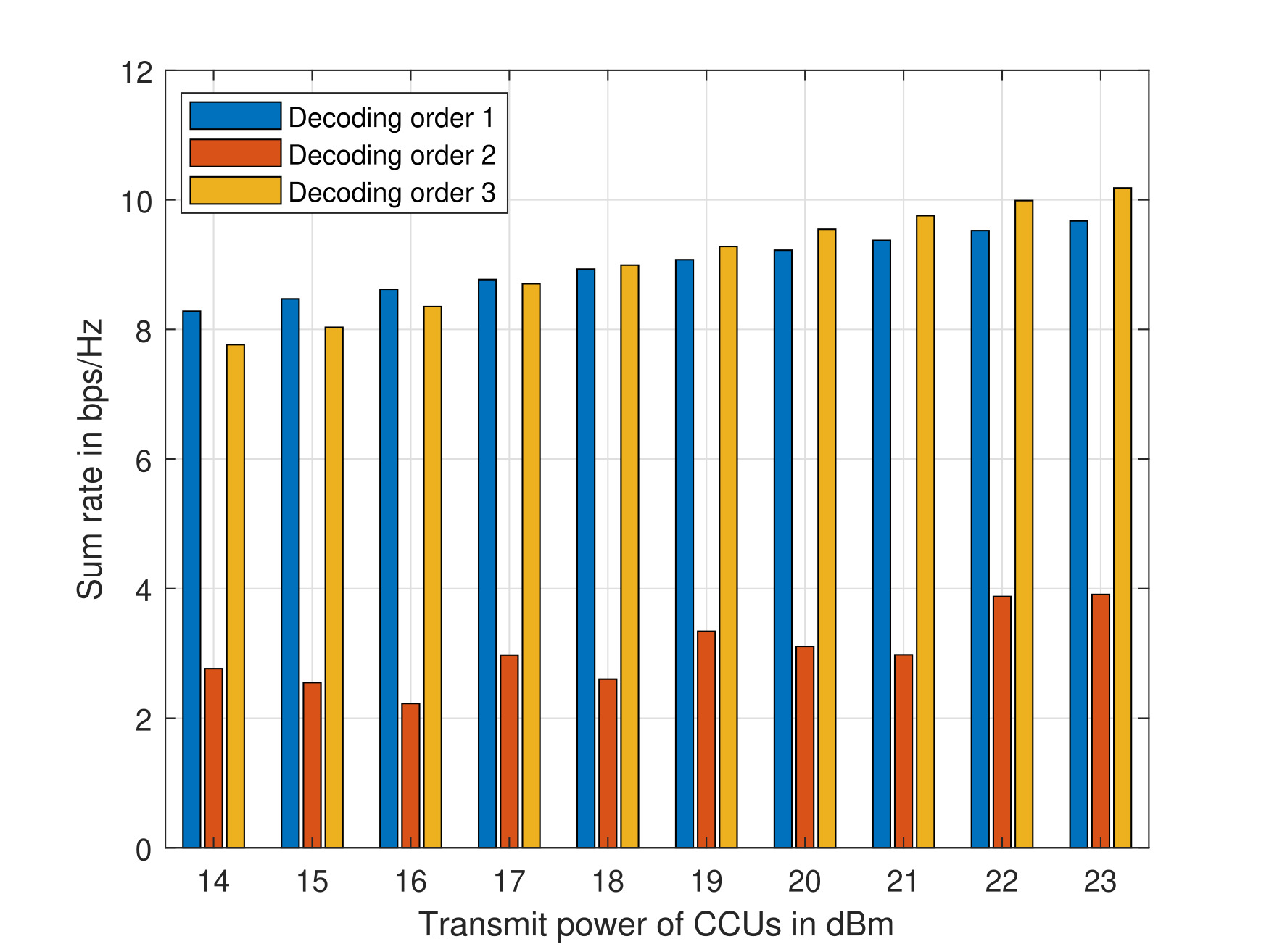}
     \caption {Sum rate vs different decoding orders while varying power budget of CEUs and $P_v^{max}=15$ dBm}
     \label{decoding_order11}
\end{figure}
{Fig. \ref{decodingorder21} depicts the impact of decoding order versus the power budget of CEUs. It can be seen from the figure that decoding order 1 and decoding order 3 achieves better performance than decoding order 2. It is because when we decode the sub-messages/messages utilizing decoding orders 1 and 3, uplink C-RSMA can leverage the benefits of flexible interference management while decoding at the BS.  On the other hand, when we decode the sub-messages of a particular user sequentially, it cannot leverage the benefits of uplink RSMA, resulting in a less achievable sum rate. Similar to Fig.\ref{decodingorder21}, it can be seen from Fig.\ref{decoding_order11} that as we increase the power budget of CCUs, the average sum rate increases, and decoding orders 1 and 3 achieve higher sum rates. With the above observations, we can conclude that it is better not to decode the sub-messages sequentially to better benefit from the C-RSMA-based approaches. Based on the above discussion, in our other simulation results, we adopted decoding order 3 as our decoding order to evaluate the impact of different parameters.
}

\begin{figure}[!t]
    \centering
\includegraphics[width=0.8\columnwidth]{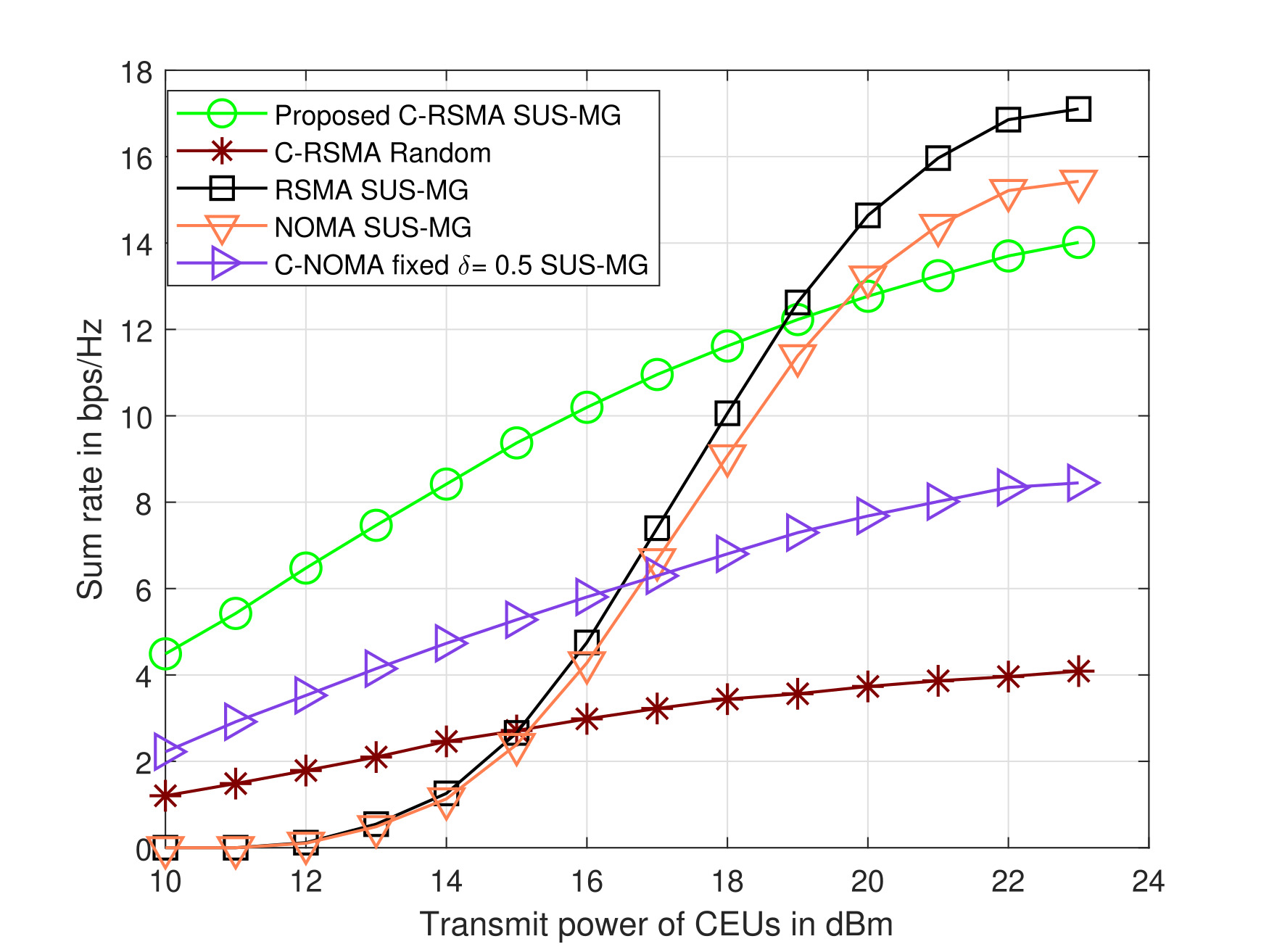}
\caption{{Sum rate vs. power budget of CEUs, $P_u^{max}=23$ dBm}}
\label{transmit_power2}
\end{figure}

\begin{figure}[!t]
    \centering
\includegraphics[width=0.8\columnwidth]{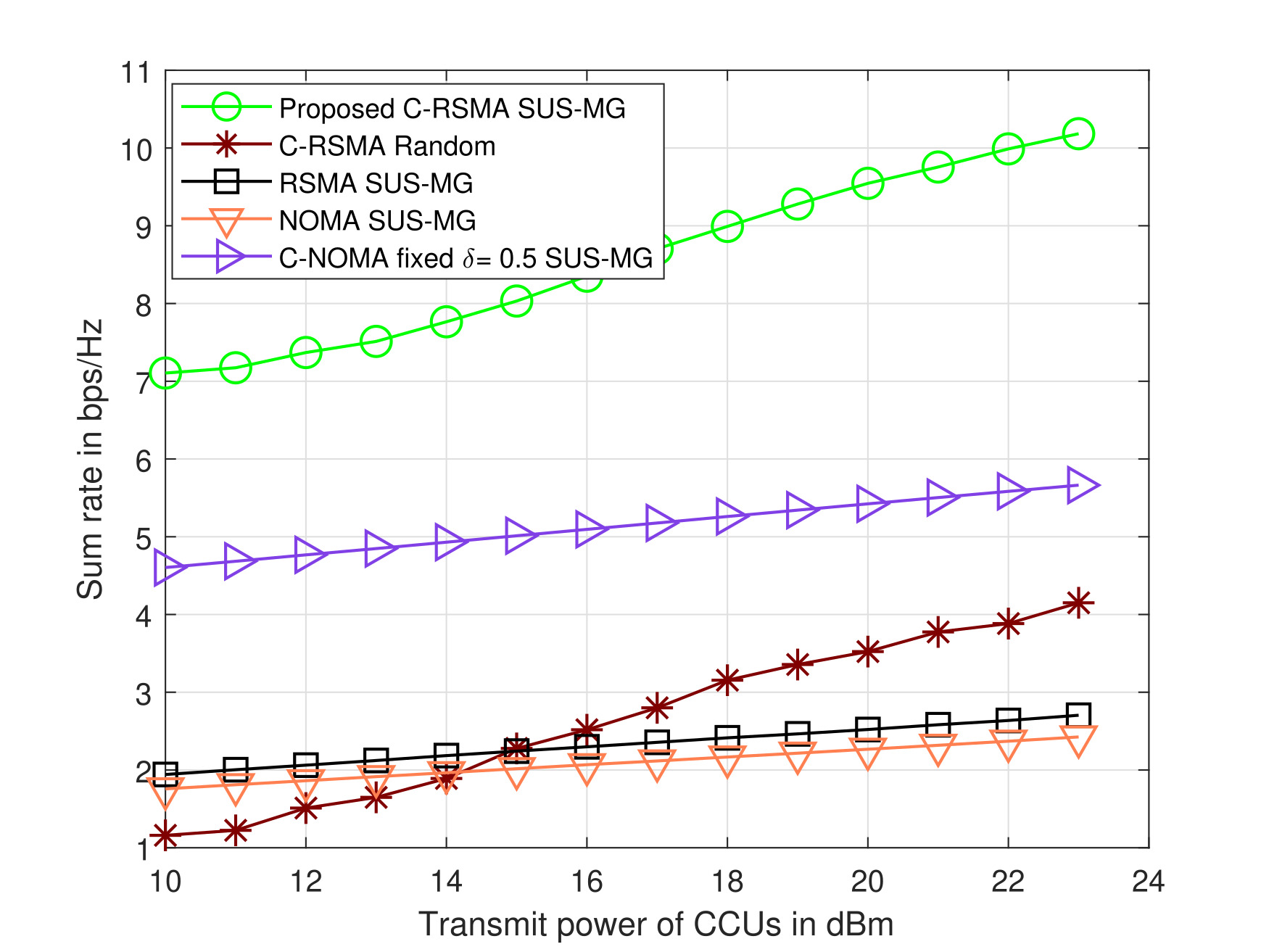}
\caption{{Sum rate vs. power budget of CCUs, $P_v^{max}=15$ dBm}}
\label{transmit_power}
\end{figure}
\begin{figure}[!t]
    \centering
\includegraphics[width=0.8\columnwidth]{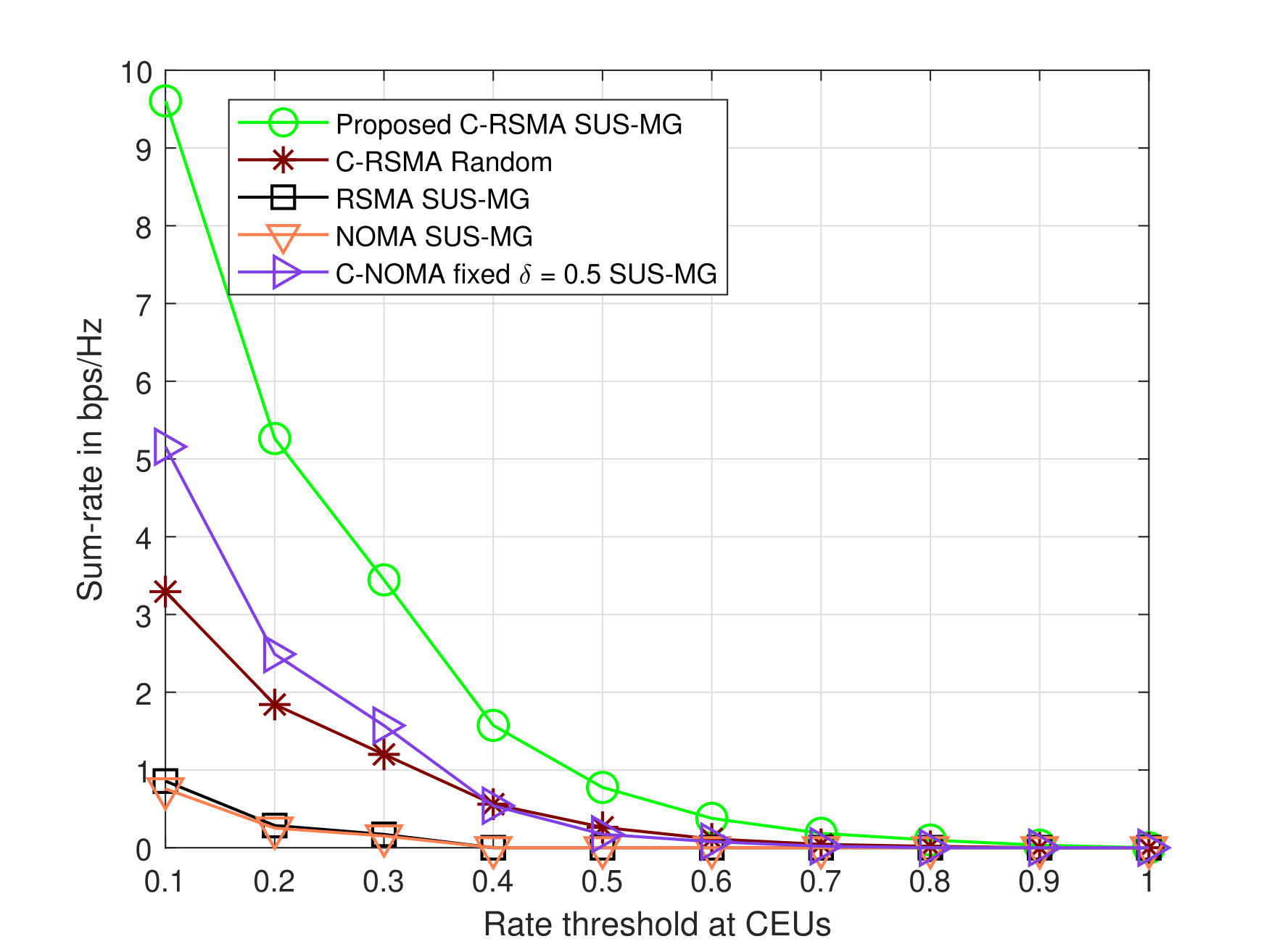}
\caption{{Sum rate vs. rate threshold of CEUs when $P_u^{max}=23$ dBm and $P_v^{max}=15$ dBm}}
\label{rate_threshold}
\end{figure}

\subsection{Impact of varying the transmit power of CEUs}
Fig. \ref{transmit_power2} represents the average sum rate of the proposed approach and the baseline schemes as we vary the power of CEU in each pair. One can see from this figure that the average sum rate of the network increases as we increase the power budget of CEU. {In the beginning, when the power budget of CEUs is around $10$ dBm, the increase of the sum rate remains modest. When the power budget starts increasing more than $15$ dBm, we can notice a significant jump in the sum rate. This is because as we increase the power budget of the CEUs, during the DT phase, the CEUs can transmit signals with more power, resulting in an increased achievable rate. Particularly when the power budget of the CEUs is high enough, it can overcome the bad channel condition with the BS, and CCUs can also achieve an improved signal quality. Our proposed scheme achieves better performance in terms of sum rate among all other schemes  until 18 dBm. This is due to the RSMA-based frameworks having the freedom to achieve better signal quality as they can play with interference levels. Meanwhile, C-RSMA with random achieves lower gain due to the random pairing schemes. On the other hand, non-cooperative RSMA achieves lower gains when the power budget of CEUs is low, and when the power budget of CEUs becomes high, it outperforms the C-RSMA scheme. This is because when the power budget of CEUs is low, they cannot overcome the poor channel condition of CEUs. However, when CEUs have enough power budget, they can overcome the bad effects of poor channel conditions, and even without cooperation, they can achieve higher gains. }
\subsection{Impact of varying the transmit power of CCUs}
Fig. \ref{transmit_power} presents the average sum rate achieved by the proposed scheme with other compared schemes versus the power budget at the CCUs in each pair. {It can be seen from Fig. \ref{transmit_power} that as we increase the transmit power from 10 dBm to 23 dBm, the average sum rate of all the strategies starts to increase. This is because as we increase the power budget of CCUs, the CCUs can help to transmit with more power during the CT phase, which helps to boost the average sum rate of both CCUs and CEUs at the BS. Meanwhile, the sum-rate of non-cooperative techniques remains modest as the power budget of CEUs is not high and hence, non-cooperative techniques cannot overcome the poor channel condition between CEUs and BS.}
\subsection{Impact of varying the rate threshold of CEU}
   {Fig. \ref{rate_threshold} demonstrates the average sum rate over the rate threshold of the CEUs for our proposed and all other schemes. As we increase the rate threshold of CEUs, the sum rate starts to decrease for all the strategies. This is because as we increase the rate threshold of the CEUs, the available power budget of the CEU is not sufficient to meet the high data rate requirements by overcoming the bad effects of the poor channel condition. However, it can be seen that the cooperative schemes achieve higher rates than the non-cooperative ones. More particularly, when the CEUs rate threshold exceeds 0.4 bps/Hz, non-cooperative schemes fail to achieve any sum rate, and the solution becomes infeasible. Meanwhile, cooperative schemes achieve better performance even at higher rate thresholds. However, in all cases, our proposed C-RSMA scheme with pairing achieves the best performance in both low- and high-rate requirements.}
\section{Conclusion}
In this paper, the problem of sum-rate maximization for the uplink C-RSMA in a multi-user scenario is investigated by jointly optimizing user pairing and power allocation at the UEs subject to the constraints of transmit power at the UEs and the required QoS in terms of the minimum achievable data rate. Due to the non-convexity of the joint optimization problem, we adopted the bi-level optimization, which decouples the problem into two sub-problems. The first sub-problem is the user pairing problem where a CCU and a CEU are paired by adopting the SUS-MG algorithm. Particularly, each CCU is selected utilizing the SUS algorithm which assures a semi-orthogonality among CCUs exists in order to reduce the interference as much as possible. Afterward, a CEU is paired with a CCU considering an MG-based algorithm where the channel gains between the users are considered as a preference utility. In the second sub-problem, the power allocation is performed per pair by invoking a SCA-based low-complexity algorithm. Our simulation results demonstrated that our proposed approach achieved the best average sum-rate compared to other conventional schemes. 
\label{Conclusion}

\bibliographystyle{IEEEtran}

\vfill

\end{document}